\documentclass[runningheads]{llncs}

\newcommand\beq{\begin{equation}}
                \newcommand\enq{\end{equation}}

\newcommand{\LC}{$LTL_{LIA}$}

\newcommand{\RightBox}{{\phantom{a}}\hfill $\Box$ \\}

\newcommand{\diamin}{\Diamond\kern-0.5em{\raisebox{.25ex}{\rm -}}\kern0.175em}

\newcommand{\ldot}{{\rm <}\kern-0.37em{\raisebox{.25ex}{\bf .}}\kern0.375em}

\newcommand{\tup}[1]{\langle#1\rangle}

\newcommand{\prefun}{{\ensuremath{\tt pre}}}
\newcommand{\postfun}{{\ensuremath{\tt post}}}

\newcommand{\xrun}[1]{\xrightarrow[]{#1}}

\newcommand{\A}{\mathcal{A}}
\newcommand{\B}{\mathcal{B}}

\newcommand{\N}{\mathcal{N}}

\newcommand{\pre}[1]{\raisebox{0.2em}{\(\bullet\)}{#1}}
\newcommand{\post}[1]{{#1}\raisebox{0.2em}{\(\bullet\)}}

\usepackage{packages}
\begin{document}

\title{Two dimensional bounded model checking for unbounded Petri nets}

\author{Ramchandra Phawade \inst{1} \and S Sheerazuddin \inst{2} \and Tephilla Prince \inst{3}}

\authorrunning{{Phawade} et al.}

\institute{ IIT Dharwad, India \email{prb@iitdh.ac.in} \and
    NIT Calicut, India \email{sheeraz@nitc.ac.in} \and
    IIT Dharwad, India \email{tephilla.prince.18@iitdh.ac.in
    }
}

\maketitle
\begin{abstract}
    Petri nets are a well-established formal model for modeling and analysis of distributed and concurrent systems. Existing verification approaches for unbounded Petri nets typically rely on interleaved executions, which may not fully capture the system's concurrent behavior and they verify only reachability and coverability properties~\cite{Abdulla04,AbdullaIN00,AbdullaJRS06,AmatDH22}. In this work, we consider unbounded Petri nets with concurrent semantics and specify its properties in LTL with linear integer arithmetic, {\LC}.

    We extend the standard BMC algorithm by placing a bound on the number of tokens in the net in each execution step. We call this new algorithm two-dimensional $2$D-BMC, since we bound both the number of tokens and the execution length simultaneously. We implement our $2$D-BMC algorithm for verifying unbounded Petri nets against {\LC} properties in our tool DCModelChecker. Further, we evaluate the performance of DCModelChecker against state-of-the-art Petri net verification tools.

\end{abstract}

\section{Introduction}
The model checking problem decides whether the runs of a given system P satisfy its specification $\alpha$. To tackle state space explosion~\cite{ClarkeKNZ11}, we employ  bounded model checking (BMC), which is model checking on \emph{bounded} runs of the system. Petri nets are a well-studied representation of distributed and concurrent systems. For their verification, their system properties are represented in some temporal logic~\cite{P77}. Semantically speaking, the execution of nets can be performed by either interleaving the transition firings (interleaving semantics) or concurrently firing all enabled transitions (true concurrent semantics). True concurrency was introduced by Petri in ``Kommunikation mit Automaten" in the context of implementing a Turing Machine, where he proposes an asynchronous architecture where storage space can be extended while the machine continues to operate. In practice, existing Petri net verification tools employ interleaving semantics~\cite{Abdulla04,AbdullaIN00,AbdullaJRS06,AmatDH22}. In~\cite{EsparzaH01}, they employ an unfolding (KenMcMillan's~\cite{KenUnfolding92}) based approach to verify LTL-X properties on bounded nets. To the best of our knowledge, there are no tools for verifying nets with true concurrent semantics for unbounded nets.

While it is impossible to extend the storage space while the machine operates, one can think of an unbounded Petri net, which allows for an apriori unknown number of tokens in the net in any run of bounded length.
With regards to model checking of unbounded nets, their safety properties were verified using the backwards reachability technique~\cite{AbdullaIN00,AbdullaJRS06}, and coverability using KenMcMillan's notion of unfolding~\cite{Abdulla04}. In~\cite{AmatDH22}, they give a property directed reachability based semi-decision procedure to check generalized reachability on generalized Petri nets using SMT. We identify the gap in verifying properties beyond reachability and coverability for unbounded nets.

In an early version of this work~\cite{arxiv2dbmc}, we discussed the preliminary $2$D-BMC algorithm for verification of nets on LTL with counting, and provided the first version of DCModelChecker~\cite{FigPrince2022}. In this work, we describe two new tools, that verify properties on nets specified in LTL with linear integer arithmetic, {\LC}, and improved benchmarking. They are, DCModelChecker 1.0~\cite{Zen2DBMC}, which is used to verify nets over {\LC} properties using interleaved semantics and the $2$D-BMC algorithm. And DCModelChecker 2.0~\cite{Zen2DBMCv2}, which we built by extending DCModelChecker1.0, to allow concurrent transition firing (not always truly concurrent). 

In this work, we make the following contributions.
First, we introduce a novel extension of BMC, called $2$D-BMC. Second, we provide the SMT encoding for unbounded Petri nets and their interleaved and concurrent executions. Third, the model checking tools for verifying temporal properties for unbounded Petri nets.

\subsection*{Organisation} In this paper, in Sec.~\ref{sec:prelims} we discuss Petri nets and their semantics. In Sec.~\ref{sec:logic}, we recall the temporal logic LTL along with integer arithmetic operations and counting, which is helpful for describing the properties of the nets. Bounded Model Checking for nets is explored in Sec.~\ref{sec:encodingLTL} where the algorithm $2$D-BMC is proposed and the SMT encoding is discussed. This is implemented in the BMC tool detailed in Sec.~\ref{sec:dcmtool} and the verification experiments are given in Sec.~\ref{sec:experiments}.

\section{Preliminaries}\label{sec:prelims}
In this section we begin with the definitions and terms used in this work.
\subsection{Petri nets}\label{sec:pn}
Petri nets are a commonly used formalism for modeling the dynamic behaviour of systems~\cite{murata89,petri1962kommunikation}. Here, we recollect the standard notions associated with nets.

\begin{definition}\label{def:pn}
    A \emph{PN} is a tuple $N=(P,T,F)$ where $P$ is a finite set of  \emph{places}, $T$ is a finite set of \emph{transitions}, $F : (P \times T) \cup (T \times P)\longrightarrow \mathbb{N}$ is the \emph{flow function}.
\end{definition}

In case of weighted arcs, we additionally have the weight function $W:F\rightarrow \mathbb{N}_0$, where $\mathbb{N}_0$ is the set of non-negative integers. For instance, for the net $N_0$ in Fig.~\ref{fig:petrinet0}, $P=\{p_0,p_1,p_2,p_3\}$, $T=\{t_0,t_1,t_2,t_3\}$ and $F$ is represented graphically by the directed arcs.

\begin{figure}[ht]
    \centering
    	\begin{tikzpicture}[scale=0.45]

 \node[transition,fill=black,minimum height=.4cm,minimum width=.4cm,label=left:{\small $t_{0}$}] (t0) at (-10,0) {};
		
\node[place,label=above:{\small $p_{0}$},draw=red!50,inner
sep=0pt,minimum height=.40cm, minimum width=.5cm] (p0) at (-8,0){$\bullet$} ;	
		
\node[transition,fill=black,minimum height=.4cm,minimum width=.4cm,label=above:{\small $t_{1}$}] (t1) at (-6,0) {};

\node[place,label=above:{\small $p_{1}$},draw=red!50,inner
sep=0pt,minimum height=.40cm, minimum width=.5cm] (p1) at (-4,2){} ;
	
\node[place,label=below:{\small $p_{2}$},draw=red!50,inner
sep=0pt,minimum height=.40cm, minimum width=.5cm] (p2) at (-4,-2){} ;

\node[transition,fill=black,minimum height=.4cm,minimum width=.4cm,label=above:{\small $t_{2}$}] (t2) at (-2,2) {};

\node[transition,fill=black,minimum height=.4cm,minimum width=.4cm,label=above:{\small $t_{3}$}] (t3) at (-2,-2) {};

\node[place,label=above:{\small $p_{3}$},draw=red!50,inner
sep=0pt,minimum height=.40cm, minimum width=.5cm] (p3) at (0,0){} ;

\draw[->] (t0) -- (p0);
\draw[->] (p0) -- (t1);
\draw[->] (t1) -- (p1);
\draw[->] (t1) -- (p2);

\draw[->] (p1) -- (t2);
\draw[->] (p2) -- (t3);

\draw[->] (t2) -- (p3);
\draw[->] (t3) -- (p3);

\draw[->] (p3) node[midway, above]{} (t0);
\end{tikzpicture}
    \caption{A simple Petri net $N_0$ with marking $\langle 1,0,0,0 \rangle$}
    \label{fig:petrinet0}
\end{figure}

\begin{definition}
    The set $\pre{t}=\{p\in P\mid F(p,t)>0\}$ is called the \emph{set of pre-places} of $t \in T$.
    The set $\post{t} =\{p\in P\mid F(t,p)>0\}$ is called the \emph{set of post-places} of $t \in T$.
\end{definition}
For instance, in Fig.~\ref{fig:petrinet0}, $\pre{t_0}=\emptyset$, $\pre{t_1}=\{p_0\}$, $\pre{t_2}=\{p_1\}$, $\pre{t_3}=\{p_2\}$.
And $\post{t_0} =\{p_0\}$, $\post{t_1} =\{p_1,p_2\}$, $\post{t_2} =\post{t_3} =\{p_3\}$.
The flow function $F$ is also described by the pre and post condition functions.

\begin{definition}
    Given a PN $N=(P,T,F)$, the \emph{pre-condition function} $\prefun_N$ and the \emph{post-condition function} $\postfun_N$ are defined as follows:
    \begin{align*}
        {\prefun}_N : T\rightarrow ( P\rightarrow \mathbb{N})  & \quad\prefun_N(t)(p)=F(p,t)  \\\quad
        {\postfun}_N : T\rightarrow ( P\rightarrow \mathbb{N}) & \quad\postfun_N(t)(p)=F(t,p)
    \end{align*}
\end{definition}

The states of the PN are the distributions of tokens in the places, described by \emph{markings}.

\begin{definition}
    A \emph{marking} $M$ of a PN $N=(P,T,F)$ is a function $M : P \rightarrow \mathbb{N}_0$, where $\mathbb{N}_0$ is the set of non-negative integers. A \emph{marked PN} is a pair $PN=(N,M_0)$ where $N$ is a PN and $M_0$ is a marking, called \emph{initial marking}.
\end{definition}

In Fig.~\ref{fig:petrinet0} the marking $\langle 1,0,0,0 \rangle$, denotes the distribution of tokens in the places.

\begin{definition}[Enabledness Rule] A transition $t$ is enabled at marking $M$
    when $\forall p\in P, M(p)\geq \prefun_N (t,p)$.
\end{definition}
The \emph{enabledness rule} is a prerequisite for firing a transition $t$. On the firing of $t$, the successor of the current marking is obtained according to its pre and post conditions, i.e., by removing $F(p,t)$ tokens from each $p \in \pre{t}$ and adding $F(t,p')$ tokens to each $p' \in \post{t}$, leaving tokens in the remaining places as they are.

\begin{definition}[Firing Rule]
    Given a marking $M$ in a Petri net, on firing an enabled transition $t$, we get the successor marking $M'$, denoted by $M \xrightarrow{t} M'$, such that:
    \begin{align*}
        \forall p\in P:
        M'(p) = M(p) - F(p,t) + F(t,p)
    \end{align*}
\end{definition}

If there are several enabled transitions at a given marking, exactly one of them is non-deterministically chosen to be fired in the step.

Given the marking in Fig.~\ref{fig:petrinet0} and on firing transition $t_1$, the marking in Fig.~\ref{fig:petrinet0} is obtained.

\begin{definition}
    The set of markings \emph{reachable} from marking $M$ in a given net $N$, denoted by $reach_N(M)$, is the smallest set of markings such that:
    \begin{itemize}
        \item $M \in reach_N(M)$ and
        \item If $M'\xrightarrow{t}M''$ for some $t\in T$ and $M' \in reach_N(M)$, then $M'' \in reach_N(M)$.
    \end{itemize}
    The \emph{reachability graph} of a PN $N$ is the directed graph $(\N,E)$ where $\N$ is the set of markings of $N$, $E\subseteq \N\times\N$, and $(M,M')\in E$ iff there is some $t\in T$ such that $M\xrightarrow{t}M'$. The set of reachable markings of a marked PN $(N,M_0)$, is $reach_N(M_0)$.
\end{definition}

Markings are naturally ordered by the coverability relation $\leq$, defined as follows.

\begin{definition}
    Given two markings $M_1$ and $M_2$, we say that a marking $M_2$ covers $M_1$, denoted by $M_1 \leq M_2$, iff $\forall p \in P, M_1(p) \leq M_2(p)$.
\end{definition}

In Petri net $N_0$ from Fig.~\ref{fig:petrinet0} suppose the marking $M_0 =\langle 0,0,0,0 \rangle$ and on firing $t_0$ we obtain $M_1=\langle 1,0,0,0 \rangle$, we say that $M_1$ covers $M_0$.

The \emph{reachability problem} is as follows: given a marked PN $(N,M_0)$ and a marking $M$, whether $M\in reach_N(M_0)$. For instance, in Fig.~\ref{fig:petrinet0}, given the initial marking, $M_0$ is $\langle 1,0,0,0 \rangle$. We know that the marking $M_1=\langle 0,0,0,2 \rangle = reach_{N_0}(M_0)$. Hence, $M_1\in reach_{N_0}(M_0)$.

It is known that the reachability problem for PN is \emph{decidable} but in non-primitive recursive time~\cite{CzerwinskiLLLM21}. The \emph{coverability problem} is as follows: given a marked PN $(N,M_0)$ and a marking $M$, whether there is a reachable marking $M'\in reach_N(M_0)$ such that $M'\leq M$. It is known that also the coverability problem is decidable and in EXPSPACE~\cite{Rackoff78}. Another problem on PN is \emph{deadlock-freeness}.

\begin{definition}[Dead Marking]
    A marking $M$ for net $N$ is \emph{dead} if transition $t$ is not enabled at $M$ for all $t\in T$. The marked PN $(N,M_0)$ is \emph{deadlock-free}, if there is no dead marking in $reach_N(M_0)$.
\end{definition}

For instance, in Fig.~\ref{fig:petrinet0}, if the net is modified such that the transition $t_0$ is not present, it would result in a deadlock after obtaining the marking $\langle 0,0,0,2\rangle$.
The \emph{deadlock-freeness} problem is as follows: given a marked PN $N$, whether $N$ is deadlock-free.
It is well-known that deadlock-freeness and reachability are recursively equivalent and, thus, deadlock-freeness is decidable but in non-primitive recursive time complexity~\cite{Hack74,ChengEP95}.

\subsection{Petri net Semantics}

In this section, we look at the formal sematics of Petri nets, in particular the interleaving and concurrent semantics. The key difference in the two approaches, is in the firing condition. We recall the firing rule:

Given an enabled transition $t$ at marking $M$, on firing $t$, we get the successor marking $M'$, denoted by $M \xrightarrow{t} M'$, such that:
\begin{align*}
    \forall p\in P:
    M'(p) = M(p) - F(p,t) + F(t,p)
\end{align*}

In interleaving semantics, there is atmost one transition that is fired in an instance or step of the occurrence sequence. Given an initial marking $M_0=\tup{0,0,0,0}$, on firing $t_0$ twice and $t_1$ once, we obtain this sequence
$\tup{0,0,0,0} \xrun{t_0}\tup{1,0,0,0}\xrun{t_0}\tup{ 2,0,0,0} \xrun{t_1}\tup{0,1,1,0} $ as depicted in Fig.~\ref{fig:petrinet1}.

\begin{figure}[ht]
    \centering
    \begin{tikzpicture}[scale=0.45]

 \node[transition,fill=black,minimum height=.4cm,minimum width=.4cm,label=left:{\small $t_{0}$}] (t0) at (-10,0) {};
		
\node[place,label=above:{\small $p_{0}$},draw=red!50,inner
sep=0pt,minimum height=.40cm, minimum width=.5cm] (p0) at (-8,0){} ;	
		
\node[transition,fill=black,minimum height=.4cm,minimum width=.4cm,label=above:{\small $t_{1}$}] (t1) at (-6,0) {};

\node[place,label=above:{\small $p_{1}$},draw=red!50,inner
sep=0pt,minimum height=.40cm, minimum width=.5cm, tokens=1] (p1) at (-4,2){$\bullet$} ;
	
\node[place,label=below:{\small $p_{2}$},draw=red!50,inner
sep=0pt,minimum height=.40cm, minimum width=.5cm, tokens=1] (p2) at (-4,-2){$\bullet$} ;

\node[transition,fill=black,minimum height=.4cm,minimum width=.4cm,label=above:{\small $t_{2}$}] (t2) at (-2,2) {};

\node[transition,fill=black,minimum height=.4cm,minimum width=.4cm,label=above:{\small $t_{3}$}] (t3) at (-2,-2) {};

\node[place,label=above:{\small $p_{3}$},draw=red!50,inner
sep=0pt,minimum height=.40cm, minimum width=.5cm] (p3) at (0,0){} ;

\draw[->] (t0) -- (p0);
\draw[->] (p0) -- (t1);
\draw[->] (t1) -- (p1);
\draw[->] (t1) -- (p2);

\draw[->] (p1) -- (t2);
\draw[->] (p2) -- (t3);

\draw[->] (t2) -- (p3);
\draw[->] (t3) -- (p3);

\draw[->] (p3) node[midway, above]{} (t0);
\end{tikzpicture}
    \caption{Petri net $N_0$ with marking $\langle 0,1,1,0 \rangle$}
    \label{fig:petrinet1}
\end{figure}

\begin{figure}[ht]
    \centering
    \begin{tikzpicture}[scale=0.45]

    \node[transition,fill=black,minimum height=.4cm,minimum width=.4cm,label=left:{\small $t_{0}$}] (t0) at (-10,0) {};

    \node[place,label=above:{\small $p_{0}$},draw=red!50,inner
    sep=0pt,minimum height=.40cm, minimum width=.5cm] (p0) at (-8,0){} ;

    \node[transition,fill=black,minimum height=.4cm,minimum width=.4cm,label=above:{\small $t_{1}$}] (t1) at (-6,0) {};

    \node[place,label=above:{\small $p_{1}$},draw=red!50,inner
    sep=0pt,minimum height=.40cm, minimum width=.5cm] (p1) at (-4,2){} ;

    \node[place,label=below:{\small $p_{2}$},draw=red!50,inner
    sep=0pt,minimum height=.40cm, minimum width=.5cm] (p2) at (-4,-2){} ;

    \node[transition,fill=black,minimum height=.4cm,minimum width=.4cm,label=above:{\small $t_{2}$}] (t2) at (-2,2) {};

    \node[transition,fill=black,minimum height=.4cm,minimum width=.4cm,label=above:{\small $t_{3}$}] (t3) at (-2,-2) {};

    \node[place,label=above:{\small $p_{3}$},draw=red!50,inner
    sep=0pt,minimum height=.40cm, minimum width=.5cm] (p3) at (0,0){$\bullet$$\bullet$} ;

    \draw[->] (t0) -- (p0);
    \draw[->] (p0) -- (t1);
    \draw[->] (t1) -- (p1);
    \draw[->] (t1) -- (p2);

    \draw[->] (p1) -- (t2);
    \draw[->] (p2) -- (t3);

    \draw[->] (t2) -- (p3);
    \draw[->] (t3) -- (p3);

    \draw[->] (p3) node[midway, above]{} (t0);
\end{tikzpicture}
    \caption{Petri net $N_0$  with marking $\langle 0,0,0,2 \rangle$}
    \label{fig:petrinet2}
\end{figure}

Now, both transitions $t_2$ and $t_3$ are enabled. In interleaving semantics, one may fire either of the two transitions, one after the other.
However, in concurrent semantics, the following sequence is also additionally possible in a single step, resulting in Fig.~\ref{fig:petrinet2}.

$\langle 0,1,1,0\rangle \xrun{t_2,t_3}\langle 0,0,0,2\rangle $

\begin{definition}[Concurrent Firing Rule]

    Given a Petri net $N$ and a marking $M$ having a set of enabled transitions $\tau \in T$, we obtain the successor marking $M'$ on concurrent firing of a subset of transitions $\tau'\subseteq \tau$ using the following rule:

    \begin{align*}
        \forall p\in \pre{\tau '}:
        M'(p) = M(p) -\sum_{t\in \tau'} F(p,t) + \sum_{t \in \tau'}F(t,p)
        \text{ if } \sum_{t\in \tau'} F(p,t) - M(p) \geq 0.
    \end{align*}
\end{definition}

The condition $\sum_{t\in \tau'} F(p,t) - M(p) \geq 0$ takes care of the sufficiency of tokens at each of the pre-places of the transitions in $\tau'$ such that the transitions can be fired. This is a particularly elegant rule in the context of representing concurrent firing of transitions in a Petri net in a SMT solver.

\begin{figure}[ht]
    \centering
    \begin{tikzpicture}[scale=0.45]

\node[place,label=left:{\small $p_{0}$},draw=red!50,inner
sep=0pt,minimum height=.40cm, minimum width=.5cm] (p0) at (-4,2){$\bullet$} ;	

\node (w0) at (-2,2.5) {$w_0$};
\node (w1) at (-2,-1.5) {$w_1$};
		
\node[place,label=left:{\small $p_{1}$},draw=red!50,minimum height=.40cm, minimum width=.5cm] (p1) at (-4,-2){$\bullet\bullet$} ;

\node[transition,fill=black,minimum height=.4cm,minimum width=.4cm,label=above:{\small $t_{0}$}] (t0) at (0,2) {};

\node[transition,fill=black,minimum height=.4cm,minimum width=.4cm,label=above:{\small $t_{1}$}] (t1) at (0,-2) {};

\node (w1) at (-2,-5) {$w_1$};

\node[transition,fill=black,minimum height=.4cm,minimum width=.4cm,label=above:{\small $t_{2}$}] (t2) at (0,-6) {};

\node[place,label=right:{\small $p_{2}$},draw=red!50,minimum height=.40cm, minimum width=.5cm] (p2) at (4,-2){} ;

\node (w1) at (2.5,-1.5) {$w_2$};

\node[place,label=right:{\small $p_{3}$},draw=red!50,minimum height=.40cm, minimum width=.5cm] (p3) at (4,-6){} ;		

\node (w1) at (2.5,-5.5) {$w_3$};

\draw[->] (p0) -- (t0);
\draw[->] (p1) -- (t0);
\draw[->] (p1) -- (t1);
\draw[->] (t1) -- (p2);
\draw[->] (p1) -- (t2);
\draw[->] (t2) -- (p3);
\end{tikzpicture}
    \caption{Concurrent Firing of Petri net $N_1$}
    \label{fig:pnconcur}
\end{figure}

In Fig.~\ref{fig:pnconcur}, $P=\{p_0,p_1,p_2,p_3\}$ and $T=\{t_0,t_1,t_2\}$. Given an initial marking $M_0=\langle 1,2,0,0\rangle$, the set of enabled transitions w.r.t $M_0$, is $\tau=\{t_0,t_1,t_2\}$. Suppose the subset of enabled transitions $\tau'=\{t_1,t_2\}$, then $\pre{\tau '}=\{p_1\}$ and the step $\langle 1,2,0,0\rangle \xrun{t_1 \& t_2}\langle 0,0,1,1\rangle $ is allowed by the concurrent firing rule stated above. However, suppose $M_1=\langle 1,1,0,0\rangle$, $\tau'=\{t_1,t_2\}$  and $\pre{\tau '}=\{p_1\}$, then the step $\langle 1,1,0,0\rangle \xrun{t_1 \& t_2} $ is not allowed by the same rule.

\emph{Remark:} Consider a PN where transitions $t_1, t_2, \cdots t_n$ are enabled and there are sufficient tokens in the pre places of $t_1, t_2, \cdots t_n$ such that these transitions may be fired.
In the context of true concurrent semantics, we fire the maximal number of enabled transitions.
Consequently, this means that if there are intermediate markings, which are reachable by an interleaved firing of a subset of the enabled transitions $t_1, t_2, \cdots t_n$, then those markings become unreachable when executing in a truly concurrent manner.

\subsection{Petri net Representation for an SMT Solver}\label{sec:pnrepresentation}

The bigger picture is to verify properties of the Petri net using an SMT solver~\cite{HandbookSMTBarrettT18}. Satisfiability of formulas with respect to a particular theory such as arrays, integer arithmetic, bit vectors is called as Satisfiability Modulo Thoery. The tools employing the decision procedures and programs to solve these satisfiability constraints are known as SMT solvers. Well known SMT solvers include Z3, CVC4~\cite{MouraB08,BarrettCDHJKRT11}. In this section, we discuss the representation of nets such that they can be added as constraints in an SMT solver, namely Z3. The Petri net representation with interleaving semantics and the concurrent semantics differ. While verifying infinite state systems,~\cite{AbdullaJ01} employed backwards reachability for proving safety properties, and in case of concurrent programs, ~\cite{AbdullaJRS06} proved liveness and termination via backwards reachability. In the unbounded Petri net setting,~\cite{AbdullaIN00} is the last known work where unfoldings (in the sense of Ken McMillan~\cite{KenUnfolding92}) are discussed from a verification point of view. In state of the art Petri net verification tools, interleaving semantics is adopted~\cite{AmatDH22}. We take an alternate perspective of this and unfold unbounded Petri nets with concurrency using the help of SMT solvers. In this thesis, by unfolding, we refer to the process of obtaining the subsequent configurations from the initial marking of the net. This is not to be confused with the notion of unfolding described by Ken McMillan in his seminal work on partial orders~\cite{KenUnfolding92}.

We outline the variables and data structures that are necessary to describe the concurrent semantics. We have a finite set of transitions, $t_0,\ldots,t_m$ described in the net, their names are in the list $tNames$. We have a finite set of places, $p_0,\ldots,p_l$ their names are in the list $pNames$. Arcs can be of either of two types: where the source is a transition and the target is a place or the source is a place and the target is a transition. The net formalism does not allow arcs between places and between transitions themselves. If they occur, the net description is erroneous, and we cannot move ahead with the unfolding.

We use the vector $iWTk$ to store the expressions with respect to $k$, to compute the incident weights of transitions. We use the vector $WTk$ to store the expressions with respect to $k$, to denote the change in weights for places.
We use two-dimensional matrices $Wt[m][l]$ and $iWt[m][l]$ to denote the net change in weights and the incidence weights (outgoing from places). The expressions for the same are stored in $WTVars$ and $iWTVars$ respectively.

We construct the two-dimensional weight matrix $Wt[row][col]$ of size $m \times l$ which contains the net weight of the arcs. Initially, all the matrix entries are initialized to zero.

For an arc from place $p_i$ to transition $t_j$ with the weight $w$, we have the matrix entry $Wt[i][j] = Wt[i][j] - w$.
For an arc from transition $t_j$ to place $p_i$ with the weight $w$, we have the matrix entry $Wt[i][j] = Wt[i][j]  + w$. Now, if there are incoming and outgoing arcs of the same weights, then the net weight $Wt[i][j]=0$. Notice that, by looking only at the Wt matrix, we may not distinguish between the case where there are no arcs to and from an element of the net. Hence, it is necessary to have a separate data structure for the same. We have a two-dimensional matrix $iWt[row][col]$ of size $m \times l$ which contains the incident weights to the transitions.
For an arc from place $p_i$ to transition $t_j$ with the weight $w$, we have the matrix entry $iWt[i][j] =Wt[i][j] - w$. The marking of the net consists of the set tokens at each place and represents the state of the net at any instance.
The initial marking of the net can be obtained from the net description and contains the number of tokens in each of the places $p_0,\ldots,p_l$ in the net. The subsequent markings may be constructed from the matrix $Wt$ and using the transition function of the net. The transition function $TF$ describes the behaviour of the net. The initial marking of the net is stored in $initial$. We introduce a method $printTFTruthTable$ that can aid to visualise the transition function $TF$ which is an expression describing the function. Most utility methods and the $2-DBMC$ algorithm are similar to that of the interleaving semantics. For experimentation, we have three different versions for constructing the Transition Function which are equivalent to each other (as verified by truth tables, experiments with $352$ properties) and are a simplification of the expression using $\land \land $, $||$ instead of $\implies$ and so on. Experiments suggest that one version is slightly faster than the others. The variable $T[ti]$ denotes the $i$th transition being fired. Hence the expression $!T[ti]$ denotes that the $i$-th transition is not fired. For every pair of transitions and places, if there are no outgoing arc from the transition $t_i$ then the expression $preCond$ containing the precondition for firing of transitions is constructed as follows:

\begingroup
\allowdisplaybreaks
\begin{align}
    if (emptyOutTi)     & \{                                                  \\
    preCond             & = (tmp == iWt[pi][ti] \lor tmp == 0)                \\
                        & \land ((T[ti] \land  tmp == iWt[pi][ti])            \\
                        & \lor (!T[ti] \land tmp == 0))                       \\
    cumulativeIncidentW & = tmp                                               \\
    emptyOutTi          & = !emptyOutTi\}                                     \\
    else\{              &                                                     \\
    preCond             & = preCond \land  (tmp == iWt[pi][ti] \lor tmp == 0) \\
                        & \land ((T[ti]\land tmp == iWt[pi][ti])              \\
                        & \lor  (!T[ti] \land tmp == 0))                      \\
    cumulativeIncidentW & = cumulativeIncidentW + tmp\}
\end{align}
\endgroup
The sum of incident weights at a transition $t_i$ is stored in $cumulativeIncidentW$ as an expression of the $iWt[pi][ti]$ if there is an arc from the transition $t_i$ to place $p_i$ or it is zero. If the transition is not fired, then there are no weights to be considered.

If the expression is non-empty, the previously constructed $preCond$ are \textbf{anded} with the newly constructed expression and the cumulative incident weights are updated in the same manner.

The post condition is constructed if the weight $Wt[pi][ti]!=0$.
\begin{align*}
    postCond          & = ((tmp == Wt[pi][ti]) \lor (tmp == 0) \\
                      & \land((T[ti] \land  tmp == Wt[pi][ti]) \\
                      & \lor (!T[ti]\land  tmp == 0))          \\
    cumulativeWChange & = tmp                                  \\
    emptyChangePi     & = !emptyChangePi
\end{align*}
Based on the above expressions, we construct the transition function $TF$
\begin{verbatim}
    if (emptyTF){
            if (!emptyOutTi){
                TF = preCond & (Px[pi] + cumulativeIncidentW >= 0) 
                emptyTF = false}
            if (!emptyChangePi){                    
                if (!emptyOutTi)
                    TF = TF & postCond 
                    &(Py[pi] == Px[pi] + cumulativeWChange)
                else
                    TF = postCond &(Py[pi] == Px[pi]
                     + cumulativeWChange)
                emptyTF = false
                }
        }
        else{
            if (!emptyOutTi)
                TF = TF &preCond 
                &((Px[pi] + cumulativeIncidentW) >= 0)
            if (!emptyChangePi)
                TF = TF &postCond
                 &(Py[pi] == (Px[pi] + cumulativeWChange))
        }
\end{verbatim}
The transition function is a conjunction of the preconditions, postconditions and the change in the markings. In case of interleaving semantics, there is an additional conjunction to the transition function, a disjunction of each transition $t_i$, to ensure that exactly one transition is fired in a step.

Now that the nets, their semantics and their representation are discussed, we shall see how to represent their properties.

\section{The Logic {\LC}}\label{sec:logic}

Linear Temporal Logic ($LTL$)~\cite{P77,VardiW86} is a natural choice to describe
the temporal properties of such systems.
We are interested in Logic {\LC} (an extension of $LTL$ with a few
differences), as it is easy to specify temporal properties as well as invariants in it.

\begin{example}\label{ex:lcprop} Recall the net $N_1$ in Fig.~\ref{fig:pnconcur}. Let $\#p_i$ be the atomic proposition denoting the token counter at place $p_i$.  Consider the invariant, where at any point in time, taken together, the places $p_2$ and $p_3$ contain exactly $0$ or $2$ tokens. This is easily expressed in {\LC} as $G((\#p_{2} + \#p_{3} =0)\lor (\#p_{2} + \#p_{3} =2))$.
\end{example}
As shown in Ex.~\ref{ex:lcprop}, the logic {\LC} is an extension of $LTL$. In the case of $LTL$, atomic formulas are propositional
constants which have no further structure.
In {\LC}, at the propositional level, we allow token counters and arithmetic operators over them.

\subsection*{Syntax}
Let $t_c$ be the set containing all token counters. The formulas in {\LC} are given as follows:
\begin{equation}
    \begin{split}
        \alpha , \hat{\alpha}&::= \#p \mid \mathfrak{c} \mid (\mathfrak{c} \ast \#p ) \mid  (\#p \ast \mathfrak{c}) \mid (\alpha + \hat{\alpha} ) \mid (\alpha - \hat{\alpha})\\ \nonumber
        \beta &::=  (\alpha < \hat{\alpha}) \mid (\alpha > \hat{\alpha}) \mid (\alpha \le \hat{\alpha}) \mid  (\alpha \ge \hat{\alpha}) \mid (\alpha=\hat{\alpha})
    \end{split}
\end{equation}

\noindent where $\#p \in t_c$ and $\mathfrak{c} \in \mathbb{Z}^+$.

\[\psi::= \beta \mid \lnot \psi  \mid \psi\lor\psi^{\prime}   \mid \psi\land\psi^{\prime}  \mid  X \psi  \mid F \psi\mid G \psi \mid \psi U  \psi^{\prime}\]
Here, the modalities X, F, G and $U$ have the same meaning as the LTL temporal operators Next, Eventually, Globally and Until, respectively.

\subsubsection*{Semantics}

The logic {\LC} is interpreted over a sequence of markings $\varrho = m_0, m_1, \ldots$ where each $m_i$ is a marking of the net. Let $CN$ be the countable set of tokens in the net.

The evaluation of the set of terms $\alpha$, denoted by $||\alpha||_i$ is defined inductively as follows:

\emph{Base case:} The term $\#p$ evaluates to a non-negative integer, $||\#p||_i$ denoting the number of tokens satisfying property $p$ at instance $i$ where $||\#p||_i=m_i(p)$.
The evaluation of $\mathfrak{c}$ at instance $i$, $||\mathfrak{c}||_i=\mathfrak{c}$.

\emph{Inductive case:}
The terms $(\mathfrak{c} \#p )$ (and $(\#p \mathfrak{c})$) evaluates to $\mathfrak{c} ||\#p||_i$ (and $||\#p||_i \mathfrak{c}$ respectively) the multiplication of an integer denoted by $\#p$ with a constant $\mathfrak{c}$.
The terms $(\alpha + \hat{\alpha})$ and $(\alpha - \hat{\alpha})$ evaluate to  addition and subtraction of integers $||\alpha||_i,||\hat{\alpha}||_i$.

The truth of a formula at an instant $i$ in the model $\varrho$ is given by the $\models$ relation defined by induction over the structure of $\psi$ as follows:

\begin{enumerate}
    \item $\varrho,i \models (\alpha < \hat {\alpha})$ iff $||\alpha||_i < ||\hat {\alpha}||_i$.

    \item $\varrho,i \models (\alpha > \hat {\alpha})$ iff $||\alpha||_i > ||\hat {\alpha}||_i$.

    \item $\varrho,i \models (\alpha \le \hat {\alpha})$ iff $||\alpha||_i \le ||\hat {\alpha}||_i$.

    \item $\varrho,i \models (\alpha \ge \hat {\alpha})$ iff $||\alpha||_i \ge ||\hat {\alpha}||_i$.

    \item $\varrho,i \models (\alpha = \hat {\alpha})$ iff $||\alpha||_i = ||\hat {\alpha}||_i$.

    \item $\varrho,i\models \lnot \psi$ iff $\varrho,i\not\models \psi$.
    \item $\varrho,i\models \psi\lor\psi^{\prime}$ iff $\varrho,i\models \psi$ or $\varrho,i\models \psi^{\prime}$.
    \item $\varrho,i\models \psi\land\psi^{\prime}$ iff $\varrho,i\models \psi$ and $\varrho,i\models \psi^{\prime}$.
    \item $\varrho,i\models X\psi$ iff $\varrho,i+1\models \psi$.
    \item $\varrho,i\models F \psi$ iff $\exists j\ge i$, $\varrho,j\models \psi$.
    \item $\varrho,i\models G \psi$ iff $\forall j\ge i$, $\varrho,j\models \psi$.
    \item $\varrho,i\models  \psi U\psi^{\prime}$ iff $\exists j\ge i$, $\varrho,j\models \psi^{\prime}$ and for all $i \le j' <j: \varrho,j'\models \psi$.

\end{enumerate}

\subsubsection{Two dimensional Bounded Semantics of {\LC}}\label{sssec:bsem}
We provide the bounded semantics~\cite{BiereCCSZ03} of {\LC} in order to arrive at the SMT encoding in Sec.~\ref{sec:encodingLTL} which is necessary for BMC.
We use $\models^{\langle\lambda,\kappa\rangle}$ as a restriction of $\models$ over bounded runs of length $\lambda$. The bound $k$ is divided into two parts: $k = \kappa + \lambda$, where $\kappa$ is a bound on the number of tokens and $\lambda$ is a bound on the execution steps of the net.
Since the runs are bounded, there are at most $\kappa$ tokens in the system, namely CN = $\{0,\ldots ,\kappa-1\}$.

\begin{figure}[ht]
    \centering
    \scalebox{0.6}{		\begin{tikzpicture}[shorten >=1pt,node distance=2cm,on grid,
		minimum width=2em, text width=2em, align=center]
		\node[state] (s)   {}; 
		\node[state,font=\Large] (si) [ right=of s] {$s_i$};
		\node[state] (si1) [ right=of si] {};
		\node[state,font=\Large] (sk) [ right=of si1] {$s_\lambda$};
		\path[->] 
		(s) edge  node {} (si)
		(si) edge  node {} (si1)
		(si1)	edge  node{} (sk);
		\end{tikzpicture}	}
    \caption{Bounded loop-free path of length $\lambda$}
    \label{fig:loopfreepath}
\end{figure}

\begin{figure}[ht]
    \centering
    \scalebox{0.6}{		\begin{tikzpicture}[shorten >=1pt,node distance=2cm,on grid,
		minimum width=2em, text width=2em, align=center]
		\node[state] (s)   {}; 
		\node[state] (sl) [ right=of s] {$s_l$};		
		\node[state] (sl1) [ right=of sl] {};
		\node[state] (si) [ right=of sl1] {$s_i$};
		\node[state] (sk) [ right=of si] {$s_\lambda$};
		\path[->] 
		(s) edge  node {} (sl)
		(sl)	edge  node{} (sl1)
		(sl1)	edge  node{} (si)
		(si)	edge  node{} (sk)
		(sk)	edge [bend right] node {}(sl)	;
		\end{tikzpicture}	}
    \caption{Bounded path with $(\lambda,l)$ - loop}
    \label{fig:klloop}
\end{figure}

First, we describe the bounded semantics without loop where $0\leq i \leq \lambda$:

\begin{enumerate}

    \item $\varrho,i \models^{\langle\lambda,\kappa\rangle} (\alpha < \hat {\alpha})$ iff $||\alpha||_i < ||\hat {\alpha}||_i$.

    \item $\varrho,i \models^{\langle\lambda,\kappa\rangle} (\alpha > \hat {\alpha})$ iff $||\alpha||_i > ||\hat {\alpha}||_i$.

    \item $\varrho,i \models^{\langle\lambda,\kappa\rangle} (\alpha \le \hat {\alpha})$ iff $||\alpha||_i \le ||\hat {\alpha}||_i$.

    \item $\varrho,i \models^{\langle\lambda,\kappa\rangle} (\alpha \ge \hat {\alpha})$ iff $||\alpha||_i \ge ||\hat {\alpha}||_i$.

    \item $\varrho,i \models^{\langle\lambda,\kappa\rangle} (\alpha = \hat {\alpha})$ iff $||\alpha||_i = ||\hat {\alpha}||_i$.

    \item $\varrho,i\models^{\langle\lambda,\kappa\rangle} \lnot \psi$ iff $\varrho,i\not\models^{\langle\lambda,\kappa\rangle} \psi$.

    \item $\varrho,i\models^{\langle\lambda,\kappa\rangle} \psi\lor\psi^{\prime}$ iff $\varrho,i\models^{\langle\lambda,\kappa\rangle} \psi$ or $\varrho,i\models^{\langle\lambda,\kappa\rangle} \psi^{\prime}$.

    \item $\varrho,i\models^{\langle\lambda,\kappa\rangle} \psi\land\psi^{\prime}$ iff $\varrho,i\models^{\langle\lambda,\kappa\rangle} \psi$ and $\varrho,i\models^{\langle\lambda,\kappa\rangle} \psi^{\prime}$.

    \item $\varrho,i\models^{\langle\lambda,\kappa\rangle} X\psi \text{ iff }
              \begin{cases}
                  \varrho,i+1\models^{\langle\lambda,\kappa\rangle} \psi     & \text{ if } (i<\lambda) \\

                  \varrho,i\not \models^{\langle\lambda,\kappa\rangle} X\psi & \text{otherwise }
              \end{cases}$

    \item[] When instance $i$ is less than the bound $\lambda$, the formula $\psi$ is evaluated at instance $i+1$ else, the formula is unsatisfiable.

    \item $\varrho,i\models^{\langle\lambda,\kappa\rangle} F \psi$ iff $\exists j: i\le j \le \lambda $, $\varrho,j\models^{\langle\lambda,\kappa\rangle} \psi$.

    \item[] This formula is satisfiable if there is some instance $j$ such that $i \le j \le\lambda$ at which the property $\psi$ holds.

    \item $\varrho,i\not \models^{\langle\lambda,\kappa\rangle} G \psi$.

    \item[] In the absence of a loop in the bounded run, the above formula is always unsatisfiable.

    \item $\varrho,i\models^{\langle\lambda,\kappa\rangle}  \psi U\psi^{\prime}$ iff $\exists j: i\le j \le \lambda$, $\varrho,j\models^{\langle\lambda,\kappa\rangle} \psi^{\prime}$ and for all $j': i \le j' <j: \varrho,j'\models^{\langle\lambda,\kappa\rangle} \psi$.

    \item[] This formula is satisfied when formula $\psi^{\prime}$ is satisfied at some instance $j$  and for all instances less than $j$, formula $\psi$ is satisfied.

\end{enumerate}

Second, we describe the bounded semantics with $(\lambda,l)$-loop~\cite{BiereCCSZ03} where $0\leq i \leq \lambda$ and $0\leq l \leq \lambda$ as in Fig.~\ref{fig:klloop}. Semantics and explanations are given only where they differ from the corresponding case without loop:

\begin{enumerate}

    \item[10.] $\varrho,i\models^{\langle\lambda,\kappa\rangle} X\psi \text{ iff }
            \begin{cases}
                \varrho,i+1\models^{\langle\lambda,\kappa\rangle} \psi & \text{ if } (i<\lambda) \\
                \varrho,l \models^{\langle\lambda,\kappa\rangle} \psi  & \text{otherwise }
            \end{cases}$

    \item[] When instance $i$ is less than the bound $\lambda$, the formula $\psi$ is evaluated at instance $i+1$ else, $\psi$ is evaluated at instance $l$, which is the next instance of $\lambda$.

    \item[11.] $\varrho,i\models^{\langle\lambda,\kappa\rangle} F \psi$ iff $\exists j: min(l,i)\le j \le \lambda$, $\varrho,j\models^{\langle\lambda,\kappa\rangle} \psi$.

    \item[] This formula is satisfiable if there is some instance $j$ such that $min(l,i) \le j \le\lambda$, where the formula $\psi$ is satisfied.

    \item[12.] $\varrho,i\models^{\langle\lambda,\kappa\rangle} G \psi$ iff $\forall j: min(l,i)\le j \le \lambda$, $\varrho,j\models^{\langle\lambda,\kappa\rangle} \psi$.

    \item[] This formula is satisfiable if for all instances $j$ such that $min(l,i) \le j \le\lambda$, the formula $\psi$ is satisfied in each instance.

    \item[13.] $\varrho,i\models^{\langle\lambda,\kappa\rangle} \psi U\psi^{\prime} \text{ iff }
            \begin{cases}
                \exists j:i \le j \le \lambda, \varrho,j\models^{\langle\lambda,\kappa\rangle} \psi^{\prime} \text{ and } & \text{ if }(i\le l) \\
                \forall j': i \le j' < j: \varrho,j' \models^{\langle\lambda,\kappa\rangle} \psi                          &                     \\
                \exists j: i \le j \le \lambda, \varrho,j\models^{\langle\lambda,\kappa\rangle}\psi^{\prime} \text{ and } & \text{ if } (i> l)  \\
                \forall j': i \le j' <j: \varrho,j' \models^{\langle\lambda,\kappa\rangle} \psi                           &                     \\
                \hspace{10em}\text{or}                                                                                    &                     \\
                \exists j: l \le j < i, \varrho,j\models^{\langle\lambda,\kappa\rangle}\psi^{\prime} \text{ and }         &                     \\
                \forall j': l \le j' < j: \varrho,j' \models^{\langle\lambda,\kappa\rangle} \psi
            \end{cases}$

    \item[] Consider the two cases:
        \begin{itemize}
            \item If $(i\le l)$, the current instance $i$ is less than or equal to the loop instance $l$, this formula is satisfied when formula $\psi^{\prime}$ is satisfied at some instance $j$  such that $i \le j \le \lambda$ and for all instances between $i$ and $j$, formula $\psi$ is satisfied.
            \item If $(i> l)$, the formula may be satisfied in either of the two intervals between $i$ to $\lambda$ or between $l$ to $i$. Hence, the formula is satisfied if either of the following are satisfied: formula $\psi^{\prime}$ is satisfied at some instance $j$  such that $i \le j \le \lambda$ and for all instances less than $j$, formula $\psi$ is satisfied or, formula $\psi^{\prime}$ is satisfied at some instance $j$  such that $l \le j < i$ and for all instances between $l$ and $j$, formula $\psi$ is satisfied.
        \end{itemize}

\end{enumerate}

Now that we have discussed the logic, in the subsequent section, we encode the bounded model checking problem for unbounded Petri nets on {\LC} using various semantics.
\section{Bounded Model Checking for unbounded Petri nets}\label{sec:encodingLTL}

In the subsequent section, we describe BMC algorithms with respect to verification of LTL properties on Petri nets. Notice that in Algorithm.~\ref{alg:bmc} and Algorithm.~\ref{alg:2dbmc}, the temporal property $\alpha$ can be replaced with properties in any temporal logic and the model can be replaced by any formal model. Where the unfolded configuration of the Petri net is mentioned (cf. Line. 11 in Algorithm.~\ref{alg:bmc} and cf. Line.14 in Algorithm.~\ref{alg:2dbmc}), the Petri net may be unfolded using either interleaving or concurrent semantics.

\begin{figure}[!ht]
    \centering

    \scalebox{0.7}{	\begin{tikzpicture}[scale=0.45]
				\tikzset{
	>=stealth',
	punkt/.style={
		rectangle,
		rounded corners,
		draw=black, very thick,
		text width=8.5em,
		minimum height=2em,
		text centered},
	pil/.style={
		->,
		thick,
		shorten <=2pt,
		shorten >=2pt,}
}
\tikzstyle{aldecision} = [diamond, draw, fill=blue!20,
text width=4.5em, text badly centered, node distance=2.5cm, inner sep=0pt]
\tikzstyle{alblock} = [rectangle, draw, fill=blue!20,
text width=5em, text centered, rounded corners, minimum height=4em]
\tikzstyle{line} = [draw, very thick, color=black!50, -latex']
\tikzstyle{allibrary} = [draw, ellipse,fill=red!20, node distance=2.5cm,
minimum height=2em]


\node[place,label=above:p0,inner
sep=0pt,minimum height=1cm, minimum width=1cm, tokens=1] (p0) at (0,0){} ;


\node[place,label=above:p1,inner
sep=0pt,minimum height=1cm, minimum width=1cm, tokens=1] (p1) at (7,0){} ;


\node[place,label=above:p2,inner
sep=0pt,minimum height=1cm, minimum width=1cm, tokens=1] (p2) at (14,0){} ;

\node[transition,minimum height=0.8cm,minimum width=0.8cm] (t5) at (-3,0) {t5};	


\node[transition,minimum height=0.8cm,minimum width=0.8cm] (t0) at (-3,-5) {t0};

\node[transition,minimum height=0.8cm,minimum width=0.8cm] (t1) at (3,-5) {t1};	

\node[transition,minimum height=0.8cm,minimum width=0.8cm] (t2) at (7,-5) {t2};	

\node[transition,minimum height=0.8cm,minimum width=0.8cm] 
(t3) at (11,-5) {t3};	

\node[transition,minimum height=0.8cm,minimum width=0.8cm] 
(t4) at (17,-5) {t4};

\node[place,label=below:p3,inner
sep=0pt,minimum height=1cm, minimum width=1cm] (p3) at (0,-10){} ;

\node[place,label=below:p4,inner
sep=0pt,minimum height=1cm, minimum width=1cm] (p4) at (7,-10){} ;

\node[place,label=below:p5,inner
sep=0pt,minimum height=1cm, minimum width=1cm] (p5) at (14,-10){} ;

\path[->] (p0) edge(t1); 
\path[->] (t0) edge(p0); 
\path[->] (t1) edge(p3);
\path[->] (p3) edge(t0);

\path[->] (t1) edge(p1); 
\path[->] (p4) edge(t1); 
\path[->] (t2) edge(p4);
\path[->] (p1) edge(t2);
\path[->] (p1) edge(t3);
\path[->] (t3) edge(p4);

\path[->] (p2) edge(t4); 
\path[->] (t4) edge(p5); 
\path[->] (p5) edge(t3);
\path[->] (t3) edge(p2);

\path[->] (t5) edge(p0);

\end{tikzpicture}}

    \caption{An unbounded net}
    \label{fig:unboundednet}
\end{figure}

\begin{figure}[!ht]
    \centering
    \scalebox{0.7}{\input{4_interleaveseq}}

    \caption{Sequence of firings via interleaving semantics}
    \label{fig:interleaveseq}
\end{figure}

\begin{figure}[!ht]
    \centering
    \scalebox{0.7}{\input{4_concurseq}}

    \caption{Sequence of firings via concurrent semantics}
    \label{fig:concurseq}
\end{figure}

\begin{example}\label{exupn}
    Given an unbounded PN Fig.~\ref{fig:unboundednet}, one possible sequence of transition firings with interleaving semantics, where exactly one enabled transition is fired at instant is shown in Fig.~\ref{fig:interleaveseq}. The below are the other possible firing sequences:
    \begin{enumerate}
        \item[]$t_2\to t_1 \to t_4 \to t_3 \to t_4$
        \item[]$t_4 \to t_2 \to t_1 \to t_2$
    \end{enumerate}

    However, with respect to Fig.~\ref{fig:unboundednet} and true concurrent semantics, we obtain the much shorter firing sequence Fig.~\ref{fig:concurseq}.

\end{example}

\begin{algorithm}[hbt!]
    \caption{The standard Bounded Model Checking algorithm for Petri nets}\label{alg:bmc}
    \textbf{Input:} Marked PN $M$, temporal property $\alpha$ in NNF and bound.\\
    \textbf{Output:} SAT with counterexample trace or UNSAT
    \begin{algorithmic}[1]
        \Require $bound \geq 0$
        \State $index \gets 0$
        \State InitializeSolver($M_{0},\alpha$)
        \If{\textsc{solver.check()}==SAT}
        \State Display trace and exit
        \EndIf
        \If{\textsc{solver.check()}==UNSAT}
        \State $index \gets 1$
        \While{$index \leq bound$}
        \State $M_{index}\gets ConstructUnfoldedNet(M,index$)
        \State InitializeSolver($M_{index}, \alpha$)
        \If{\textsc{solver.check()}==SAT}
        \State Display trace and exit
        \ElsIf{\textsc{solver.check()}==UNSAT}
        \State $index \gets index+1$
        \EndIf
        \EndWhile
        \EndIf

    \end{algorithmic}
\end{algorithm}
\subsection{Discussion on the BMC algorithms}\label{ssec:bmcdsc}
First we disambiguate the notion of bounds. In the BMC approach, the number of steps upto which the system is verified, say $k$, indirectly places a bound on the number of tokens which are present in the net, thereby restricting the number of agents that are represented in the unbounded net. For instance, when unrolling the system for $k$ steps in the BMC approach, it indirectly places a cap that there can be atmost $k$ newly generated tokens at a place (taking into consideration the weight of the arc and the initial marking). It is to be noted that the bound in BMC is not fixed apriori, which differentiates it from the parametric verification approach where the parameter (the number of clients) is fixed apriori.

The standard BMC approach for Petri nets is given in Algorithm.~\ref{alg:bmc}. The inputs for Algorithm.~\ref{alg:bmc} are the marked PN $M$ with initial marking $M_0$, temporal property $\alpha$ in Negation Normal Form (NNF) and the BMC bound. The variable $index$ keeps track of the steps for which the system is being unfolded.  Line $1$-$4$ initializes the solver and checks whether the input formula is satisfied. If it is satisfied, the counterexample trace is displayed and the procedure terminates. In the case that the formula is unsatisfiable then we unfold the net for one more step (i.e., $M_{index}$) and intialize the solver to check if the formula is satisfied or not. In the case that the formula is not satisfiable, we repeat the process of unfolding the net until we either hit the BMC bound or we obtain a counterexample trace.

\begin{algorithm}[hbt!]
    \caption{The $2D$-BMC algorithm for Petri nets}\label{alg:2dbmc}
    \textbf{Input:} Marked PN $M$, temporal property $\alpha$ in NNF and bound.\\
    \textbf{Output:} SAT with counterexample trace or UNSAT
    \begin{algorithmic}[1]
        \Require $bound \geq 0$
        \State $index \gets 0$
        \State InitializeSolver($M_{0},\alpha$)
        \If{\textsc{solver.check()}==SAT}
        \State Display trace and exit
        \EndIf
        \If{\textsc{solver.check()}==UNSAT}
        \State $index \gets 1$
        \While{$index \leq bound$}
        \For{$\lambda \gets 1$ to $bound$}
        \For{$\kappa \gets 1$ to $bound$}
        \State $M_{index}\gets ConstructUnfoldedNet(M,index,\lambda,\kappa$)
        \State InitializeSolver($M_{index}, \alpha$)
        \If{\textsc{solver.check()}==SAT}
        \State Display trace and exit
        \ElsIf{\textsc{solver.check()}==UNSAT}
        \State $index \gets index+1$
        \EndIf
        \EndFor
        \EndFor
        \EndWhile
        \EndIf

    \end{algorithmic}
\end{algorithm}

Recall that we described the two dimensional bounded semantics of {\LC} in Sec.~\ref{sssec:bsem} where $\kappa$ is a bound on the number of clients (agents) and $\lambda$ is a bound on the execution steps of the net. In Algorithm.~\ref{alg:bmc} the unfolding of the constructed net used only the $index$. In contrast, in Algorithm.~\ref{alg:2dbmc}, the unfolding of the constructed net uses variations of $\lambda$ and $\kappa$.

We illustrate the usage of the algorithms via an example.

\begin{example}
    Consider the Fig.~\ref{fig:unboundednet} with the following temporal properties.
    \begin{enumerate}
        \item First consider the $LTL$ reachability property.
              $F(p_0=0 \& p_1=0 \& p_2 = 0 \& p_3 = 1 \& p_4 = 1 \& p_5=1)$.
              It is beneficial to use Algorithm.~\ref{alg:bmc} since it is a reachability property and with $5$ steps with interleaved unfolding (or $3$ steps in case of concurrent unfolding), the marking can be reached. It is not beneficial to use Algorithm.~\ref{alg:2dbmc} here, since it will not converge to a solution faster.

        \item Second, consider the {\LC} property, where

              $F((\#x) p_3(x) > p_0(x) \land p_4(x) > p_1(x) \land p_5(x) > p_2(x))$.
              This property requires searching by varying both the number of tokens as well as the time instance, hence Algorithm.~\ref{alg:2dbmc} is suitable here.
    \end{enumerate}
\end{example}

\subsection{Encoding $2$D-BMC for nets on {\LC}}
In order to perform $2$D-BMC, we must encode the net and the property into SMT formulas. First, we describe the notation that we have used. Let $[\mathcal{M}]$ be the SMT encoding of the Petri net model of the system $\mathcal{M}$. Let $\phi$ be the {\LC} property that we want to verify. As usual we negate the property $\phi$ and let $\psi = \lnot \phi$. We also assume that $\psi$ is used in its negation normal form. The $2D$-BMC encoding of the system $\mathcal{M}$ against $\psi$ for the bound $k=\lambda + \kappa$ (where $k \ge 0$) is denoted by $[\mathcal{M},\psi]_{\langle\lambda,\kappa\rangle}$ and defined as follows:
\[[\mathcal{M},\psi]_{\langle\lambda,\kappa\rangle} = [\mathcal{M}]_{\langle\lambda,\kappa\rangle} \land \Big(\big(\lnot L_{\langle\lambda,\kappa\rangle} \land [\psi]^0_{\langle\lambda,\kappa\rangle}\big) \lor\bigvee\limits_{l=0}^k(\prescript{}{l}L_{\langle\lambda,\kappa\rangle} \land \prescript{}{l}[\psi]^0_{\langle\lambda,\kappa\rangle})\Big)\]

The bound $k$ has two parts $\lambda$ and $\kappa$: $\lambda$ gives the bound for time instances and $\kappa$ gives the bound for the number of tokens.
The SMT formula $[\mathcal{M}]_{\langle\lambda,\kappa\rangle}$ encodes the runs of $\mathcal{M}$ of bound $k=\lambda + \kappa$. The formulae $[\mathcal{M}]_{\langle\lambda,\kappa\rangle}$, for $0 \leq \lambda,\kappa \le k$ are defined later in this section.

The formulas $\prescript{}{l}L_{\langle\lambda,\kappa\rangle}$ ($0 \le l \le \lambda$) and $L_{\langle\lambda,\kappa\rangle}$ are loop conditions that are mentioned in the encoding above. For any $0 \le l \le \lambda$, $\prescript{}{l}L_{\langle\lambda,\kappa\rangle}=\mathcal{T}(s_{\lambda},s_l)$ where $s_l$ is the $l$th state in the run of $\mathcal{M}$ and $s_{\lambda}$ is the $\lambda$th state. Here, the state is defined in terms of value of the marking vector $M$. So, for any instance $i$, $s_{i}$ corresponds to the state of vector $M$ at $i$. Clearly, when $\prescript{}{l}L_{\langle\lambda,\kappa\rangle}$ holds it means there is a transition from $s_{\lambda}$ to $s_l$ which denotes a back loop to the $l$th state.  The other loop condition is defined as follows: $L_{\langle\lambda,\kappa\rangle} = \underset{0 \le l \le \lambda}{\bigvee}\prescript{}{l}L_{\langle\lambda,\kappa\rangle}$. When $L_{\langle\lambda,\kappa\rangle}$ holds, it means there is a back loop to some state in the bounded run of $\mathcal{M}$.

Formula $[\psi]^0_{\langle\lambda,\kappa\rangle}$ is the SMT
encoding of $\psi$ for the bound $k=\lambda+\kappa$ when $\psi$ is
asserted at initial instance $i=0$ and there is no loop in the run of
$\mathcal{M}$. On the other hand,
$\prescript{}{l}[\psi]^0_{\langle\lambda,\kappa\rangle}$ is the
SMT encoding of $\psi$ for the bound $k=\lambda+\kappa$ when
$\psi$ is asserted at initial instance $i=0$ and there is a back loop
to the $l$th state in the run of $\mathcal{M}$. The formula $[\psi]^i_{\langle\lambda,\kappa\rangle}$ denotes the SMT encoding of $\psi$ where the bounded run does not have a loop. The formula $\prescript{}{l}[\psi]^i_{\langle\lambda,\kappa\rangle}$ denotes the SMT encoding of $\psi$ where the bounded run contains a loop for any $i$. These encodings are extensions of similar mappings defined in~\cite{BiereTACAS99}.

The bound $k=\lambda + \kappa$ starts from $0$ and is incremented by $1$ in each (macro-)step. For a fixed $k$, $\lambda$ may start from $0$, incremented by $1$ in each (micro-)step till $k$. Simultaneously, $\kappa$ may move from $k$ to $0$ and decrement by $1$ in each (micro-)step. We look at each (micro-)step. Let the variables being used in the Boolean encoding of Petri net be $t_0,t_1,\ldots,t_{n_t}$ and  $p_0,p_1,\ldots,p_{n_p}$ to represent transitions and places respectively. We use the $i$-th copy of the transition variable for instance $i$ as follows $t_{0i},\ldots,t_{n_ti}$, place variables $p_{0j},\ldots,p_{n_pi}$, where $0 \le i \le \lambda$, in  $[\mathcal{M}]_{\langle\lambda,\kappa\rangle}$.
For any $\kappa \ge 0$, we define
$$[\mathcal{M}]_{\langle 0,\kappa\rangle} = I(s_0)\land (\bigwedge\limits_{0 \le j\le n_p}p_{j0} \le \kappa).$$ Inductively, for any $\lambda >0$,
$$[\mathcal{M}]_{\langle\lambda,\kappa\rangle} = [\mathcal{M}]_{\langle\lambda-1,\kappa\rangle} \land (T(s_{\lambda-1},s_{\lambda}) \land (\bigwedge\limits_{0 \le j\le n_p}p_{j\lambda} \le \kappa)).$$

Now that we have formally described the notations for encoding of the system $\mathcal{M}$, we derive the SMT encoding of {\LC} for $\mathcal{M}$ from the bounded semantics described earlier. We make use of concurrent semantics for unfolding the net.

\subsection*{SMT Encoding for {\LC} properties}

In this section, we describe the SMT encoding, which is necessary for implementing a bounded model checker tool for Petri nets using {\LC} specifications.
We need to introduce counter variables for each place $p$ in the input Petri net in order to define the SMT encodings of the property $\psi$.
These extra variables are as follows: $\{c_p^i \mid i \ge 0,~\mbox{and } p ~\mbox{is a place in the Petri net}\}$. Now we describe the SMT encodings of $\psi$.

We define $[\psi]^i_{\langle\lambda,\kappa\rangle}$  in an inductive manner.

\emph{Base case:} The place counter at place $p_i$ evaluates to $c_p^i$. We also allow for multiplication by the constant in a straightforward manner.
\begin{enumerate}

    \item $\prescript{}{}[\#p]^i_{\langle\lambda,\kappa\rangle} \equiv c_p^i$

    \item $\prescript{}{}[\mathfrak{c}]^i_{\langle\lambda,\kappa\rangle} \equiv \mathfrak{c}$

    \item $\prescript{}{}[\mathfrak{c} \ast \#p]^i_{\langle\lambda,\kappa\rangle} \equiv \mathfrak{c} \times c_p^i$
    \item $\prescript{}{}[\#p \ast \mathfrak{c}]^i_{\langle\lambda,\kappa\rangle} \equiv c_p^i \times \mathfrak{c}$

\end{enumerate}

\emph{Induction case:} The encodings of the formulas in $\beta$ are given as follows:
\begin{enumerate}[resume*]
    \item $\prescript{}{}[\alpha + \hat{\alpha} ]^i_{\langle\lambda,\kappa\rangle} \equiv \prescript{}{}[\alpha ]^i_{\langle\lambda,\kappa\rangle} + \prescript{}{}[\hat{\alpha}]^i_{\langle\lambda,\kappa\rangle}$

    \item $\prescript{}{}[\alpha - \hat{\alpha} ]^i_{\langle\lambda,\kappa\rangle} \equiv \prescript{}{}[\alpha ]^i_{\langle\lambda,\kappa\rangle} - \prescript{}{}[\hat{\alpha}]^i_{\langle\lambda,\kappa\rangle}$

    \item[]

    \item $\prescript{}{}[\alpha < \hat{\alpha}]^i_{\langle\lambda,\kappa\rangle} \equiv [\alpha]^i_{\langle\lambda,\kappa\rangle} < \prescript{}{}[\hat{\alpha}]^i_{\langle\lambda,\kappa\rangle}$

    \item $\prescript{}{}[\alpha > \hat{\alpha}]^i_{\langle\lambda,\kappa\rangle} \equiv [\alpha]^i_{\langle\lambda,\kappa\rangle} > \prescript{}{}[\hat{\alpha}]^i_{\langle\lambda,\kappa\rangle}$

    \item $\prescript{}{}[\alpha \le \hat{\alpha}]^i_{\langle\lambda,\kappa\rangle} \equiv [\alpha]^i_{\langle\lambda,\kappa\rangle} \le \prescript{}{}[\hat{\alpha}]^i_{\langle\lambda,\kappa\rangle}$

    \item $\prescript{}{}[\alpha \ge \hat{\alpha}]^i_{\langle\lambda,\kappa\rangle} \equiv [\alpha]^i_{\langle\lambda,\kappa\rangle} \ge \prescript{}{}[\hat{\alpha}]^i_{\langle\lambda,\kappa\rangle}$

    \item $\prescript{}{}[\alpha = \hat{\alpha}]^i_{\langle\lambda,\kappa\rangle} \equiv [\alpha]^i_{\langle\lambda,\kappa\rangle} = \prescript{}{}[\hat{\alpha}]^i_{\langle\lambda,\kappa\rangle}$

    \item[]

    \item $\prescript{}{}[q]^i_{\langle\lambda,\kappa\rangle} \equiv q_i$

    \item $\prescript{}{}[\lnot q]^i_{\langle\lambda,\kappa\rangle} \equiv \lnot q_i$

    \item $\prescript{}{}[\psi \lor \psi^{\prime}]^i_{\langle\lambda,\kappa\rangle} \equiv \prescript{}{}[\psi]^i_{\langle\lambda,\kappa\rangle} \lor \prescript{}{}[\psi^{\prime}]^i_{\langle\lambda,\kappa\rangle}$

    \item $\prescript{}{}[\psi \land \psi^{\prime}]^i_{\langle\lambda,\kappa\rangle} \equiv \prescript{}{}[\psi]^i_{\langle\lambda,\kappa\rangle} \land \prescript{}{}[\psi^{\prime}]^i_{\langle\lambda,\kappa\rangle}$

    \item $\prescript{}{}[X \psi]^i_{\langle\lambda,\kappa\rangle} \equiv
              \begin{cases} \prescript{}{}[\psi]^{i+1}_{\langle\lambda,\kappa\rangle} & \mbox{if $i<\lambda$}\\ \mbox{False} & \mbox{otherwise}\end{cases}$

    \item $\prescript{}{}[F \psi]^i_{\langle\lambda,\kappa\rangle} \equiv \bigvee\limits_{i \le j \le \lambda}[\psi]^j_{\langle\lambda,\kappa\rangle}$

    \item $\prescript{}{}[G \psi]^i_{\langle\lambda,\kappa\rangle} \equiv \mbox{False}$

    \item $\prescript{}{}[\psi U\psi^{\prime}]^i_{\langle\lambda,\kappa\rangle} \equiv \underset{i\le j \le \lambda}{\bigvee}([\psi^{\prime}]^{j}_{\langle\lambda,\kappa\rangle} \land \underset{i\le j' < j}{\bigwedge}[\psi]^{j'}_{\langle\lambda,\kappa\rangle})$

\end{enumerate}

For cases allowing for a loop in the subformula, we define $\prescript{}{l}[\psi]^i_{\langle\lambda,\kappa\rangle}$ inductively, where $l$ denotes the loop index. We build on the semantics in Section.~\ref{sssec:bsem} and define the encoding of $\psi$ with respect to $l$ inductively:
\begin{enumerate}
    \item[16.] $\prescript{}{l}[X \psi]^i_{\langle\lambda,\kappa\rangle} \equiv
            \begin{cases} \prescript{}{l}[\psi]^{i+1}_{\langle\lambda,\kappa\rangle} & \mbox{if $(i<\lambda)$}\\ \prescript{}{l}[\psi]^l_{\langle\lambda,\kappa\rangle} & \mbox{if $(i=\lambda)$}\end{cases}$

    \item[17.] $\prescript{}{l}[F\psi]^i_{\langle\lambda,\kappa\rangle} \equiv \underset{min(l,i)\le j \le \lambda}{\bigvee}\prescript{}{l}[\psi]^j_{\langle\lambda,\kappa\rangle}$

    \item[18.] $\prescript{}{l}[G \psi]^i_{\langle\lambda,\kappa\rangle} \equiv \underset{min(l,i)\le j \le \lambda}{\bigwedge}\prescript{}{l}[\psi]^j_{\langle\lambda,\kappa\rangle}$

    \item[19.] $\prescript{}{l}[\psi\mathbf{U}\psi^{\prime}]^i_{\langle\lambda,\kappa\rangle} \equiv
            \begin{cases}
                \underset{i\le j \le \lambda}{\bigvee}(\prescript{}{l}[\psi^{\prime}]^{j}_{\langle\lambda,\kappa\rangle} \land \underset{i\le j' < j}{\bigwedge}\prescript{}{l}[\psi]^{j'}_{\langle\lambda,\kappa\rangle})                  & \mbox{if $(i\le l)$} \\
                                                                                                                                                                                                                                            &                      \\
                \bigg( \big(\underset{i\le j \le \lambda}{\bigvee}(\prescript{}{l}[\psi^{\prime}]^{j}_{\langle\lambda,\kappa\rangle} \land \underset{i\le j' < j}{\bigwedge}\prescript{}{l}[\psi]^{j'}_{\langle\lambda,\kappa\rangle})\big) &                      \\
                \hspace{25mm}\text{or}                                                                                                                                                                                                      & \mbox{if $(i> l)$}   \\
                \big(\underset{l\le j < i}{\bigvee}(\prescript{}{l}[\psi^{\prime}]^{j}_{\langle\lambda,\kappa\rangle} \land \underset{l\le j' < j}{\bigwedge}\prescript{}{l}[\psi]^{j'}_{\langle\lambda,\kappa\rangle}) \big) \bigg)        &                      \\
            \end{cases}$

\end{enumerate}

We have successfully provided the SMT encoding of the logic {\LC}. Now, we shall see how to encode the $2$D-BMC problem.

\subsection*{Encoding of $2$D-BMC for nets }

In order to determine the encoding of $[M]_{\langle 0,\kappa\rangle}$, we need to perform \textbf{and} operation on the initial configuration and the place constraints. The place constraints are due to the $\kappa$ variable, for each of the places.
$$[M]_{\langle 0,\kappa\rangle} = I(s_0)\land (\bigwedge_{0 \le j\le n_p}p_{j0} \le \kappa).$$

The variable with respect to time instances is denoted as $\lambda$. Hence, it is also a positive integer. In order to determine $[M]_{\langle\lambda,\kappa\rangle}$, we can make use of the previously encoded $[M]_{\langle\lambda-1,\kappa\rangle}$ and unfold the transition relation by one length, and \textbf{and} it with the place constraints as done previously (cf. Fig.\ref{2dbmcunfolding}). Hence we get the resultant relation:
$$[M]_{\langle\lambda,\kappa\rangle} = [M]_{\langle\lambda-1,\kappa\rangle} \land (T(s_{\lambda-1},s_{\lambda}) \land (\bigwedge_{0 \le j\le n_p}p_{j\lambda} \le \kappa)).$$

The encoding of the  $\neg, \land, \lor$ operators are straightforward. We also define the encoding using the temporal operators.

Notice that $\prescript{}{}[(\#x>k)p(x)]^i_{\langle\lambda,\kappa\rangle}$ is encoded using the counter $c_p$ at the place $p$, and this is translated as $ c_p > \kappa$. Similarly, we have

$\prescript{}{}[(\#x>k)p(x)]^i_{\langle\lambda,\kappa\rangle} \equiv c_p > \kappa$.  This encoding is similar for the loop-free formula as well.

\begin{figure}[!ht]
    \centering
    \scalebox{0.7}{	\begin{tikzpicture}[>=stealth,scale=0.9] 
			\tikzset{
	>=stealth',
	punkt/.style={
		rectangle,
		rounded corners,
		draw=black, very thick,
		text width=8.5em,
		minimum height=2em,
		text centered},
	pil/.style={
		->,
		thick,
		shorten <=2pt,
		shorten >=2pt,}
}
\tikzstyle{aldecision} = [diamond, draw, fill=blue!20, text width=4.5em, text badly centered, node distance=2.5cm, inner sep=0pt]
\tikzstyle{alblock} = [rectangle, draw, fill=blue!20, text width=5em, text centered, rounded corners, minimum height=4em]
\tikzstyle{line} = [draw, very thick, color=black!50, -latex']
\tikzstyle{allibrary} = [draw, ellipse,fill=red!20, node distance=2.5cm, minimum height=2em]

	\foreach \i in {0,2.5,5,7.5,9.5} {
		\draw[gray,very thin] (0,\i) -- (9.5,\i);
		\draw[gray,very thin] (\i,0) -- (\i,9.5);
	}
	\draw[->] (-.5,4) -- (-.5,5) node[above] {$\lambda$};
	\draw[->] (4.5,-0.5) -- (5.5,-0.5) node[right] {$\kappa $};
	
	\begin{scope}[every node/.style={
		circle,draw,
		minimum size=.8cm,
		align=center,
		font=\footnotesize,
		inner sep=0pt
	}]
	
	\node[rectangle] at (1.2,1.3) (00) {$[\mathcal{M},\psi]_{\langle 0,0\rangle}$}; 
	\node[rectangle] at (1.2,3.8) (10) {$[\mathcal{M},\psi]_{\langle 1,0\rangle}$}; 
	\node[rectangle] at (1.2,6.1) (20) {$[\mathcal{M},\psi]_{\langle 2,0\rangle}$};
	\node[rectangle,draw=none,rotate=90] at (1.2,8.5) (30) {$\cdots$};
	\node[rectangle] at (3.6,1.3) (01) {$[\mathcal{M},\psi]_{\langle 0,1\rangle}$};
	\node[rectangle] at (3.6,3.8) (11) {$[\mathcal{M},\psi]_{\langle 1,1\rangle}$};
	\node[rectangle] at (6.3,1.3) (02) {$[\mathcal{M},\psi]_{\langle0,2\rangle}$};
	
	\draw[red,->] (00) -- (01);
	\draw[red,->] (01) -- (10);
	\draw[blue,->] (10) -- (02);
	\draw[blue,->] (02) -- (11);	
	\draw[blue,->] (11) -- (20);  
	\draw[->] (20) -- (30); 


	\end{scope}
	
	\end{tikzpicture}}
    \caption{Unfolding of the encoded formula  $[\mathcal{M},\psi]_{(\lambda,\kappa)}$ with respect to $\lambda$ (execution steps) and $\kappa$ (number of tokens)}

    \label{2dbmcunfolding}
\end{figure}

\subsubsection{Unfolding the encoded formula}

The unfolding of the formula $[\mathcal{M},\psi]_{\langle\lambda,\kappa\rangle}$ for each $k$ with respect to $\lambda$ (execution steps) and $\kappa$ (number of tokens), upto $k=2$ is depicted pictorially in Fig.~\ref{2dbmcunfolding}. Initially, when $k=\lambda+\kappa=0$, $k=0$ (bound), $\lambda=0$ (time instance), $\kappa=0$ (number of tokens), the formula $[\mathcal{M},\psi]_{\langle 0,0\rangle}$ is evaluated. If this is found to be satisfiable, then, a witness is obtained, and we are done. If not, in the next macro-step of the unfolding, where $k=\lambda+\kappa=1$, there are possibly two micro-steps to be explored $\langle\lambda,\kappa\rangle=\langle 0,1\rangle$ and $\langle\lambda,\kappa\rangle=\langle 1,0\rangle$. First, $\kappa$ is incremented and the formula $[\mathcal{M},\psi]_{\langle 0,1\rangle}$ is evaluated. If this found to be unsatisfiable, $\lambda$ is incremented and the formula $[\mathcal{M},\psi]_{\langle 1,0\rangle}$ is evaluated. In each micro-step of the unfolding, either $\lambda$ or $\kappa$ are incremented, and the resulting formula is verified. The Fig.~\ref{2dbmcunfolding} represents one among many possible ways of exploring the state space in the system. Uisng the encoding in this section, we build the model checking tool.

\section{DCModelchecker : A tool to verify Unbounded PNs}\label{sec:dcmtool}
We build DCModelChecker 2.0 for verification of unbounded nets against {\LC} as well as LTL properties using $2$D-BMC. The previous version,  DCModelChecker 1.0~\cite{Zen2DBMC} can verify counting properties with interleaving semantics. We incrementally extended DCModelChecker 1.0, adding support for concurrent semantics to the tool, to obtain DCModelChecker 2.0~\cite{Zen2DBMCv2} which supports both. In the rest of this section, when we mention tool, we refer to DCModelChecker 2.0. We give the architecture and the overview of the tool in Section~\ref{sec:arch}. We describe the workflow in Section~\ref{sec:workflow}, and in Section~\ref{sec:experiments}, we report the experiments. The experiments are easily reproducible by directly executing the scripts in our artifact~\cite{Zen2DBMCv2}.

\subsection{Architecture}\label{sec:arch}

The general architecture of the tool is shown in Fig.~\ref{fig:dc2arch}. The tool has two primary inputs- system description in standard PNML format and the property to be tested expressed in {\LC}. The system description using Petri nets in PNML format can be obtained from the vast collection of industrial and academic benchmarks available at MCC~\cite{MCC} or created using a Petri net Editor~\cite{PNWolfgang,BerthomieuV06}. We make use of both types of benchmarks, for comparative testing. While several Petri net verification tools perform bounded model checking,
ours is unique in the model checking strategy, displaying
the counterexample and useful in particular for verifying temporal properties and invariants of unbounded nets.

\begin{figure}[!ht]
    \begin{minipage}{0.5\textwidth}
        \centering

        \scalebox{0.5}{\tikzset{
	>=stealth',
	punkt/.style={
		rectangle,
		rounded corners,
		draw=black, very thick,
		text width=8.5em,
		minimum height=2em,
		text centered},
	pil/.style={
		->,
		thick,
		shorten <=2pt,
		shorten >=2pt,}
}
\tikzstyle{aldecision} = [diamond, draw, fill=blue!20,
text width=4.5em, text badly centered, node distance=2.5cm, inner sep=0pt]
\tikzstyle{alblock} = [rectangle, draw, fill=blue!20,
text width=5em, text centered, rounded corners, minimum height=4em]
\tikzstyle{line} = [draw, very thick, color=black!50, -latex']
\tikzstyle{allibrary} = [draw, ellipse,fill=red!20, node distance=2.5cm,
minimum height=2em]	
\begin{tikzpicture}[node distance = 5em, auto, fit label/.style={yshift={(height("#1")+4pt)/2},
inner ysep={(height("#1")+8pt)/2}, label={[anchor=north,font=\itshape]north:#1}}]
	
	\node [ node distance=2em, xshift=-3em,fill=green!20] (property) {Property Formula};
	\node [ node distance=2em, below of=property, yshift=-1em, fill=green!20] (sysdesc) {System Description};
	\node [alblock, node distance=7em, text width=7em,right of=property,yshift=-1em, xshift=4em] (ppm) {Pre-Processing\\ Module};

	\node [alblock, node distance=7em, right of=ppm, xshift=1em, minimum width=4em] (tool) {$2D$- BMC Module};

	\node [allibrary, right of= tool,xshift=1em,rotate=90](z3){Z3 Solver};

	\node [ node distance=2em, below of=tool,yshift=-2em,,xshift=-2em, fill=green!20] (sat) {sat + trace};
	\node [ node distance=2em, right of=sat,  xshift=3em,fill=green!20] (unsat) {unsat};

	\path [line] (sysdesc) -- (ppm);
	\path [line] (property) -- (ppm);

	\path [line] (ppm) -- (tool);

	\path [line] (z3) -- (tool);
	\path [line] (tool) -- (z3);
	\path [line] (tool) -- (sat);
	\path [line] (tool) -- (unsat);
	\path [line] (unsat) --([xshift=2em]unsat.east) -- ([xshift=1em,yshift=-1em]tool.east) -- (tool);

\end{tikzpicture}}
        \caption{DCModelChecker  2.0 architecture}
        \label{fig:dc2arch}

    \end{minipage}
    \begin{minipage}{0.5\textwidth}
        \centering

        \scalebox{0.5}{\tikzset{
	>=stealth',
	punkt/.style={
		rectangle,
		rounded corners,
		draw=black, very thick,
		text width=8.5em,
		minimum height=2em,
		text centered},
	pil/.style={
		->,
		thick,
		shorten <=2pt,
		shorten >=2pt,}
}
\tikzstyle{aldecision} = [diamond, draw, fill=blue!20,
text width=4.5em, text badly centered, node distance=2.5cm, inner sep=0pt]
\tikzstyle{alblock} = [rectangle, draw, fill=blue!20,
text width=5em, text centered, rounded corners, minimum height=4em]
\tikzstyle{line} = [draw, very thick, color=black!50, -latex']
\tikzstyle{allibrary} = [draw, ellipse,fill=red!20, node distance=2.5cm,
minimum height=2em]	
\begin{tikzpicture}[node distance = 5em, auto, fit label/.style={yshift={(height("#1")+4pt)/2},
inner ysep={(height("#1")+8pt)/2}, label={[anchor=north,font=\itshape]north:#1}}]
	
		\node [ node distance=2em,  yshift=-1em, fill=green!20,text width=5em] (inputexpr) {Input \\Expression};
	
	\node [ node distance=2em, below of=inputexpr,yshift=-4em,fill=green!20] (grammar) {Grammar};

	\node [alblock, node distance=5em, text width=5em,right of=inputexpr, xshift=4em, minimum width=5em] (lexer) {Lexer +\\Parser};
	\node [alblock, node distance=4em, below of=lexer, yshift=-2em, text width = 7em, minimum height=3em] (antlrtool) {ANTLR tool};
	\node [alblock, node distance=4em, right of=lexer,xshift=4em] (parsetree) {parse tree};

	\begin{scope}
	\node [draw=black!50, minimum width=10em, fit={(lexer)  ([yshift=1em]lexer.north)},
	label={[anchor=north,font=\itshape]north:Language Recognizer}] (lr) {};

	\end{scope}

	\node [alblock, node distance=3.5em, right of=parsetree,xshift=4em, text width=5em] (listener) {Listener\\Walker};
	
	
	\node [ node distance=3em,  text width = 4 em, right of=listener,xshift=4em, fill=green!20] (ptree) {output tree};
	
	\node [allibrary,node distance=12em, above of= parsetree,yshift=-5em,xshift=1em](arl){ANTLR Runtime Library};
		
	\node[fit=(antlrtool)(lr)(parsetree)(listener),draw,minimum height=14em, label={[anchor=south,font=\itshape]south:Pre-Processing Module}] (ppm){};

	\path [line] (inputexpr) -- (lexer);
	\path [line] (lexer) -- (parsetree);
	\path [line] (parsetree) -- (listener);
	\path [line] (listener) -- (ptree);
	\path [line,dashed] (arl)--(lr);
	\path [line,dashed] (lr)--(arl);
	\path [line,dashed] (arl)--(listener);
	\path [line,dashed] (listener)--(arl);

	\path [line] (grammar) -- (antlrtool);
	\path [line] (antlrtool) -- (lr);
	\path [line] (antlrtool.east) -| (listener.south);
	

\end{tikzpicture}}
        \caption{Pre-Processing the model and formula using ANTLR}
        \label{ANTLRflow}

    \end{minipage}
\end{figure}

First, the two inputs are fed to the pre-processing module and consequently to the DCModelChecker 2.0 tool. The objective of the pre-processing module is to read and validate the two inputs.
We make use of ANTLR~\cite{ANTLRParrF11} to achieve this.
We give the grammar of the nets and properties to the tool so that it can recognize it against its respective grammar as shown in
Fig.~\ref{ANTLRflow}. ANTLR generates a parser for that language that can automatically build parse trees representing how a grammar matches the input.
The parse trees can be walked to construct the required data structures.
We have hand coded the grammar for the PNML format of both
types and the grammar of {\LC}, to be used by ANTLR. This is explained in detail in Sec.~\ref{sec:preprocess}.

The tool reads the output of the pre-processing module and checks if the model satisfies the property or not. We make use of the Z3 SAT/SMT Solver~\cite{MouraB08}, to solve the encoded formula and give us a result of unsatisfiable, or satisfiable with a counterexample trace. If unsatisfiable, the tool can increment the bound and look further, until the external termination bound is hit, according to the BMC algorithm. We chose Z3, for its wide industrial applications, developer community support, and ease of use. The detailed workflow is discussed in the subsequent section.

\subsection{Workflow}\label{sec:workflow}

\begin{figure}[!ht]
    \begin{minipage}{0.35\textwidth}
        \centering

        \scalebox{0.6}{\tikzset{
	>=stealth',
	punkt/.style={
		rectangle,
		rounded corners,
		draw=black, very thick,
		text width=8.5em,
		minimum height=2em,
		text centered},
	pil/.style={
		->,
		thick,
		shorten <=2pt,
		shorten >=2pt,}
}
\tikzstyle{aldecision} = [diamond, draw, fill=blue!20,
text width=4.5em, text badly centered, node distance=2.5cm, inner sep=0pt]
\tikzstyle{alblock} = [rectangle, draw, fill=blue!20,
text width=5em, text centered, rounded corners, minimum height=4em]
\tikzstyle{line} = [draw, very thick, color=black!50, -latex']
\tikzstyle{allibrary} = [draw, ellipse,fill=red!20, node distance=2.5cm,
minimum height=2em]		
		
\begin{tikzpicture}[node distance = 5em, auto]
\node [node distance=5em,text width = 5em,fill=green!20] (negate) {Negated property $\psi$};
\node [left of=negate,text width=8em,xshift=-4em,text width = 5em,fill=green!20] (readip) {System Description (M) and bound};

\node [aldecision,below of=negate,text width=10em, node distance=14em] (decide0) {At $k=0$,\\ $\lambda=0,~\kappa=0$\\ Is $[\mathcal{M},\psi]_{(0,0)}$ SAT?};
\node [allibrary, right of=negate, ,xshift=1em] (Z3k0) {Z3};
\node [text width=8em,left of=decide0,xshift=-8em,text width = 8em,fill=green!20] (counterex0) {Counter example found at $k=0$. EXIT};
\node [allibrary, below of=decide0, node distance=12em,fill=green!20] (nocounterex0) {A};
			
\path [line] (readip) -- (decide0);
\path [line] (negate) -- (decide0);
\path [line,dashed] (Z3k0) -- (decide0);
\path [line,dashed] (decide0) -- (Z3k0);
			
\path [line] (decide0) -- node [near start, color=black] {Yes} (counterex0);
			
\path [line] (decide0) -- node [, color=black,text width=10em,left] {No. Counter example is not found at $k=0$}(nocounterex0);		
			
\end{tikzpicture}}
        \caption{Workflow for $k=0$}
        \label{fig:work0}

    \end{minipage}\hfill
    \begin{minipage}{0.65\textwidth}
        \centering
        \scalebox{0.6}{\tikzset{
	>=stealth',
	punkt/.style={
		rectangle,
		rounded corners,
		draw=black, very thick,
		text width=8.5em,
		minimum height=2em,
		text centered},
	pil/.style={
		->,
		thick,
		shorten <=2pt,
		shorten >=2pt,}
}
\tikzstyle{aldecision} = [diamond, draw, fill=blue!20,
text width=4.5em, text badly centered, node distance=2.5cm, inner sep=0pt]
\tikzstyle{alblock} = [rectangle, draw, fill=blue!20,
text width=5em, text centered, rounded corners, minimum height=4em]
\tikzstyle{line} = [draw, very thick, color=black!50, -latex']
\tikzstyle{allibrary} = [draw, ellipse,fill=red!20, node distance=2.5cm,
minimum height=2em]		
\begin{tikzpicture}[node distance = 5em, auto]
\node [allibrary, below of=decide0, node distance=12em,fill=green!20] (nocounterex0) {A};
\node [alblock,text width=12em,below of=nocounterex0] (k1) {If  $k <$ bound \\Assign $k=1$\\Else EXIT};
\node [allibrary, right of=k1,xshift=6em] (Z3k0) {Z3};
\node [aldecision,below of=k1,text width=10em, node distance=12em] (decidek) {At $k=1$,\\$\lambda = 0,\kappa = 1 $\\ Is $[\mathcal{M},\psi]_{(0,1)}$ SAT?};
\node [left of=decidek,xshift=-8em,text width = 8em,fill=green!20] (sat10) {Counter example found at $k=1$ $\lambda = 0,~\kappa = 1$. EXIT};

\path [line] (nocounterex0) -- (k1);
\path [line] (k1) -- (decidek);
\path [line,dashed] (Z3k0) |- (decidek);
\path [line,dashed] (decidek) -| (Z3k0);
		
\path [line] (decidek) -- (sat10);
\path [line] (decidek) -- node [near start, color=black] {Yes} (sat10);

\node [aldecision,below of=decidek,text width=10em, node distance=18em] (decidek0l1) {At $k=1$,\\$\lambda = 1,\kappa = 0 $\\ Is $[\mathcal{M},\psi]_{(1,0)}$ SAT?};
\path [line] (decidek) -- node [left,near start, color=black,text width=14em] {No. For $k=1$, explore other $\lambda,\kappa$ values.} (decidek0l1);
\path [line,dashed] (Z3k0) |- (decidek0l1);
\path [line,dashed] (decidek0l1) -| (Z3k0);
		
\node [left of=decidek0l1,xshift=-8em,text width = 8em,fill=green!20] (satk0l1) {Counter example found at $k=1$ $\lambda = 1, \kappa = 0$. EXIT};
		
\path [line] (decidek0l1) -- (satk0l1);
\path [line] (decidek0l1) -- node [near start, color=black] {Yes} (satk0l1);

\node [alblock, below of =decidek0l1, text width=8em, yshift=-6em,node distance = 8em] (k2) {If  $k <$ bound\\Assign $k=2$\\And Continue\\Else EXIT};
\path [line] (decidek0l1) -- node [left,near start, color=black,text width=2em] {No} (k2);
\node [allibrary, right of=k2, node distance=10em,fill=green!20] (B) {B};	
\path [line] (k2) -- (B);
\end{tikzpicture}}
        \caption{Workflow for $k=1$}
        \label{fig:work1}

    \end{minipage}
\end{figure}

The system description $M$, the property formula $\phi$, and external termination bound $k$ are given to us. First, we negate the property, $\psi=\neg \phi$. Note that the negation normal form of $\psi$ is always used in the BMC process. For the bound $k=0$ and the corresponding micro-step
$\langle\lambda,\kappa\rangle=\langle 0,0\rangle$, we construct the formula $[M,\psi]_{\langle 0,0\rangle}$ and feed it to the solver. If the above formula is satisfiable, the property $\phi$ is violated in the initial configuration and we have a {\bf witness} at $k=0$ and we can stop our search. If the base case is unsatisfiable, we consider the next micro-step $\langle \lambda,\kappa\rangle$ -- with the bound $k=\lambda+\kappa=1$ -- and
construct the formula $[M,\psi]_{\langle \lambda,\kappa\rangle}$
and feed it to the solver. If this formula is satisfiable, the property
$\phi$ is violated for $k=\lambda+\kappa$ and we have a {\bf witness} for
this $k$ and we can stop our search. Otherwise, we may continue with the
next micro-step $\langle \lambda',\kappa'\rangle$ and so on. The order in
which the micro-steps are considered is illustrated in Fig.~\ref{2dbmcunfolding}. We have depicted the workflow only up to $k=1$ in Fig.~\ref{fig:work1}. In practice, for the property $\phi$, we search until we have considered all micro-steps for the termination
bound $k$. As expected, for the given model and bound $100$ (i.e, when parameter $k$ reaches $100$) it is observed that a counterexample is not found, hence we terminate. This means that the property holds for all traces of the model up to the length $100$.

\subsection{Pre-processing Petri nets}\label{sec:preprocess}

In order to build a verification tool that uses Petri nets, we require a standard way of representation. While, various Petri net verification tools exist, recently, the most widely accepted standard is the Petri net Markup Language (PNML)~\cite{HillahKPT10}. PNML is based on the Extensible Markup Language, or XML for short, which lends itself to interoperability between tools, while ensuring readability. We employ the ANTLR~\cite{ANTLRParrF11} tool for parsing the PNML input. 

This tool is compatible with any Petri net verification/analysis tools that use the PNML format. Additionally, one may also use existing graphical tools to visualise the Petri nets, limited only by the size of the Petri net~\cite{PNWolfgang}. This decision was made early in the work, to enable reuse of the prototypes and modules built in the tool chain by other research groups and to ensure reproducibility of results.We list down the actual \textbf{lexer} and \textbf{parser} grammars that were used to pre-process the nets in the following section.

\subsection*{Lexer rules: }
First, we name the following set of rules, such that we can refer to in the parsing phase:

\begin{lstlisting}
lexer grammar PNMLLexer;
\end{lstlisting}

Consider the following rule that recognizes the escape sequences for tab spaces, carriage return and new line respectively and skips over them when they occur in the input file (PNML file).

\begin{lstlisting}
S           :   [ \t\r\n]+               -> skip ;
\end{lstlisting}

The PNML format is a type of markup file which contains tags. We have the following rules to identify the open, close and comment tags; the digits and text:
\begin{lstlisting}[multicols=2]
COMMENT     :   '<!--' .*? '-->'  -> skip;
SPECIAL_OPEN:   '<?'          -> pushMode(INSIDE) ;
OPEN        :   '<'           -> pushMode(INSIDE) ;
DIGIT       :   [0-9]+ ;
TEXT        :   ~[<&]+ ; //16 bit char except < and &
mode INSIDE;
CLOSE       :   '>'          -> popMode ;
SPECIAL_CLOSE:  '?>'         -> popMode ;
SLASH_CLOSE :   '/>'         -> popMode ;
\end{lstlisting}
We have rules to tokenize the keywords and special characters used in the PNML file:
\begin{lstlisting}[multicols=2]
SLASH       :   '/' ;
EQUALS      :   '=' ;
QUOTE       :   '"' ;
SQUOTE      :   '\'' ;
PLACE       :   'place' ;
TRANSITION  :   'transition' ;
ARC         :   'arc';
INITIAL     :   'initialMarking' ;
INSCRIPTION :   'inscription' ;
TEXTTAG     :   'text' ;
USCORE      :   '_' ;
SOURCE      :   'source' ;
TARGET      :   'target' ;
ID          :   'id' ;
STRING      :   '"' ~[<"]* '"'
            |   '\'' ~[<']* '\'';
\end{lstlisting}
We have a special set of rules for the name tag. The name consists of an alphabet and may be succeded by any combination of digits, alphabets and special symbols (hyphen, underscore, dot), as allowed by the Petri net graphical analysis tool that we used~\cite{PNWolfgang}.
\begin{lstlisting}
Name        :   NameStartChar NameChar* ;
NameChar    :   NameStartChar
            |   '-' | '_' | '.' | DIGIT
            |   '\u00B7'
            |   '\u0300'..'\u036F'
            |   '\u203F'..'\u2040';
NameStartChar  :   [:a-zA-Z]
            |   '\u2070'..'\u218F'
            |   '\u2C00'..'\u2FEF'
            |   '\u3001'..'\uD7FF'
            |   '\uF900'..'\uFDCF'
            |   '\uFDF0'..'\uFFFD';
\end{lstlisting}

\subsection*{Parser rules: }

Now that we have the lexer rules that can identify and tokenize the input, in this section, we write the the parser rules which will be used by ANTLR to construct the parse tree from the valid input and throw errors if any.

First, we set the specific vocabulary of the parser as \textbf{PNMLLexer} (the set of lexer rules we defined above):

\begin{lstlisting}
parser grammar PNMLParser;
options {
    tokenVocab = PNMLLexer;
}
\end{lstlisting}

A valid PNML file contains a header followed by valid elements:

\begin{lstlisting}
doc: header element;
\end{lstlisting}
An element may be either an open tag followed by a close tag or an empty tag, each containing its name and possibly containing a set of attributes. There may be open tags for exactly one of the predefined keywords such as place, transition etc. Notice that it is not possible to define just a close tag using these set of rules. If such an input exists, our tool parses the PNML file and invalidates it by throwing a suitable error.
\begin{lstlisting}
header: '<?' Name attribute* '?>';
element:
       '<' Name attribute* '>' (
        place
        | transition
        | arc
        | element
        | textTagDigit
        | textTag
        | TEXT
        )* '<' '/' Name '>'
        | '<' Name attribute* '/>';
\end{lstlisting}

The place tag contains a mandatory identifier attribute and may contain details to represent the initial marking or special text for parsing.
\begin{lstlisting}
place: '<' 'place' 'id' '=' STRING '>' (initial | element | TEXT)* '<' '/' 'place' '>';
\end{lstlisting}

The rule for parsing the initial marking:
\begin{lstlisting}
initial:
    '<' INITIAL '>' (textTagDigit | textTag | element | TEXT)* '<' '/' INITIAL '>';
\end{lstlisting}

The rule for parsing transitions, with a mandatory identifier.
\begin{lstlisting}
transition:
    '<' 'transition' 'id' '=' STRING '>' (element | TEXT)* '<' '/' 'transition' '>';
\end{lstlisting}
The rule to recognize arcs from a place/transition to transition/place along with their identifiers.
\begin{lstlisting}
arc:
    '<' 'arc' 'id' '=' STRING source target '>' (
        inscription
        | element
        | TEXT
    )* '<' '/' 'arc' '>'
    | '<' 'arc' 'id' '=' STRING source target '/>';
source: 'source' '=' STRING;
target: 'target' '=' STRING;
\end{lstlisting}
Some additional rules to parse the text and specify the type of net for validation:
\begin{lstlisting}
inscription:
    '<' INSCRIPTION '>' (textTagDigit | textTag | element | TEXT)* '<' '/' INSCRIPTION '>';
textTagDigit: '<' TEXTTAG '>' DIGIT '<' '/' TEXTTAG '>';
textTag: '<' TEXTTAG '>' TEXT '<' '/' TEXTTAG '>';
attribute: ('id' | Name) '=' STRING;
\end{lstlisting}

Similarly, the rules are written to validate the logic {\LC} as well and are available in our artifact. As illustrated, in Fig.~\ref{fig:dc2arch}, given a valid system description in a PNML file and a valid specification  which are validated by the pre-processing module, the $2D$-BMC algorithm verifies the specification using Z3 and produces a counterexample trace or times out. In the next section, we look at the experiments conducted on our bounded model checking tool.

\section{Benchmarks and Comparisons}\label{sec:experiments}
In this section, we detail the experimental evaluation of DCModelChecker 2.0. The goals of our evaluation are to demonstrate:
\begin{enumerate}
    \item  the correctness of results returned by DCModelChecker 2.0 in comparison with the state-of-the-art ITS-Tools~\cite{ITSMCC},
    \item  the comparison of DCModelChecker 2.0 on FireabilityCardinality properties which are not verifiable by tools in the MCC for various bounds,
    \item verification of unbounded Petri nets in comparison with other tools,
    \item the strength of DCModelChecker 2.0 is in additionally verifying invariants specified in {\LC} which are not verifiable by tools in the MCC nor DCModelChecker 1.0.
\end{enumerate}

To replicate the experiments and plots, a collection of simple shell and python scripts are publicly available in our artifact~\cite{Zen2DBMCv2}.

\noindent\textbf{Benchmarks:}
In the first set of experiments described in Section~\ref{exp:compits} we consider benchmark models from the MCC and translated the LTLFireability properties into {\LC}. In our second set of experiments  in Section~\ref{exp:compv1v2}, we make use of the MCC benchmark models and write the properties containing both fireability and cardinality constraints for those models, which are not expressible in the language of MCC. In the third set of experiments, we make use of synthetic unbounded Petri nets~\cite{AmatDH22}. The fourth experiment demonstrates verification of invariants using our own synthetic benchmark given in the artifact.

\subsection{Experiment 1: Computing the Tool Confidence and Verification of LTLFireability properties - DCModelChecker 2.0 vs the state-of-the-art tool}\label{exp:compits}

To compute the Tool Confidence,  22 benchmarks from the MCC were verified using DCModelChecker 2.0 and compared against the state-of-the-art model checking tool ITS-Tools~\cite{ITSMCC}. ITS-Tools has 100 \% tool confidence~\cite{CompResultMCC22}.
The comparative experiments were performed with a timeout of $3600$ seconds.

\noindent\textbf{Evaluation Criteria:}
ITS-Tools returns True when the property holds true and False when the property does not hold. DCModelChecker 2.0 however, returns a satisfying counterexample to the negation of the property when the property does not hold. When DCModelChecker 2.0 returns UNSAT, it means that the property holds up to the given bound, and that the property may become false at a greater bound. Hence, for comparison, we may categorize the results of both tools as shown in Table~\ref{table:categoryresults}.\begin{table}[]
    \centering
    \resizebox{0.7\columnwidth}{!}{

        \begin{tabular}{p{10pt}c|cc|}
            \cline{3-4}
                                                                                                      &                                                                    & \multicolumn{2}{c|}{\begin{tabular}[c]{@{}c@{}}ITS Tools\end{tabular}}                                                                                                                                                                        \\ \cline{3-4}
                                                                                                      &                                                                    & \multicolumn{1}{c|}{\begin{tabular}[c]{@{}c@{}}True\end{tabular}}                                                          & \begin{tabular}[c]{@{}c@{}}False\end{tabular}                                                                    \\ \hline
            \multicolumn{1}{|c|}{\multirow{2}{*}[18pt]{\rotatebox[origin=c]{90}{DCModelChecker 2.0}}} & \begin{tabular}[c]{@{}c@{}}SAT\\ Property is False\end{tabular}    & \multicolumn{1}{c|}{\begin{tabular}[c]{@{}c@{}}\\A\\ Erroneous output by\\ DCModelChecker \\$\mid A \mid =7$\end{tabular}} & \begin{tabular}[c]{@{}c@{}}\\B\\ Both tools agree\\ on the property being false \\$\mid B \mid =85$\end{tabular} \\[26pt] \cline{2-4}
            \multicolumn{1}{|c|}{}                                                                    & \begin{tabular}[c]{@{}c@{}}\\UNSAT\\ Property is True\end{tabular} & \multicolumn{1}{c|}{\begin{tabular}[c]{@{}c@{}}C \\$\mid C \mid =72$\end{tabular}}                                         & \begin{tabular}[c]{@{}c@{}}D \\$\mid D \mid =179$\end{tabular}                                                   \\ \hline
        \end{tabular}
    }
    \caption{Categorization of LTLFireability results}
    \label{table:categoryresults}

\end{table}

The instances in category A are those false positive properties where DCModelChecker 2.0 returned a counterexample, whereas ITS-Tools returned false. The instances in category B are those where both tools agreed that the property was false, this is particularly significant since it gives the correctness of DCModelChecker 2.0. The instances in category C are those where DCModelChecker 2.0 returns UNSAT and ITS-Tools returns true. The instances in category D are those where DCModelChecker 2.0 returns UNSAT and ITS-Tools returns false. Those instances in categories C and D may become false at a greater bound, hence we do not conclusively take them into account.
We consider only the properties that fall in category A and B for computing the confidence ratio. This is similar to the MCC tool confidence computation~\cite{RulesMCC22} which excludes those values that are not computed by the tool. The Confidence of a tool is a value $C_{tool} \in [0,1]$, where 1 denotes that the tool returns the correct result that is agreed upon by atleast 3 participating tools and 0 denotes that the tool never returns the commonly agreed result (otherwise referred to as trusted values). In the LTL Formulas examination category, ITS-Tools has a tool confidence of 100\%,  $C_{ITS-Tools}= 1$. Hence, it is sufficient to compare DCModelChecker 2. 0 against ITS-Tools to check for the correctness of the results. The $C_{tool}$ is computed as follows:

\begin{equation}
    C_{tool} = \frac{\mid V_{tool}\mid}{\mid V \mid}
\end{equation}

\noindent where $\mid V\mid $ is the number of trusted values, which we obtain from ITS-Tools. $\mid V_{tool} \mid$ is the number of results obtained by DCModelChecker 2.0 within the set of trusted values, such that $V \subseteq V_{tool}$. For Experiment 1, comparing LTL Fireability properties of DCModelChecker and ITS-Tools, we obtained the following tool confidence $C_{DCModelChecker}= \frac{\mid A \mid}{\mid A + B \mid} = 0.90$. The underlying $2D$-BMC strategy is not complete.

\begin{figure}[ht]
    \centering
    \scalebox{0.4}{\includegraphics{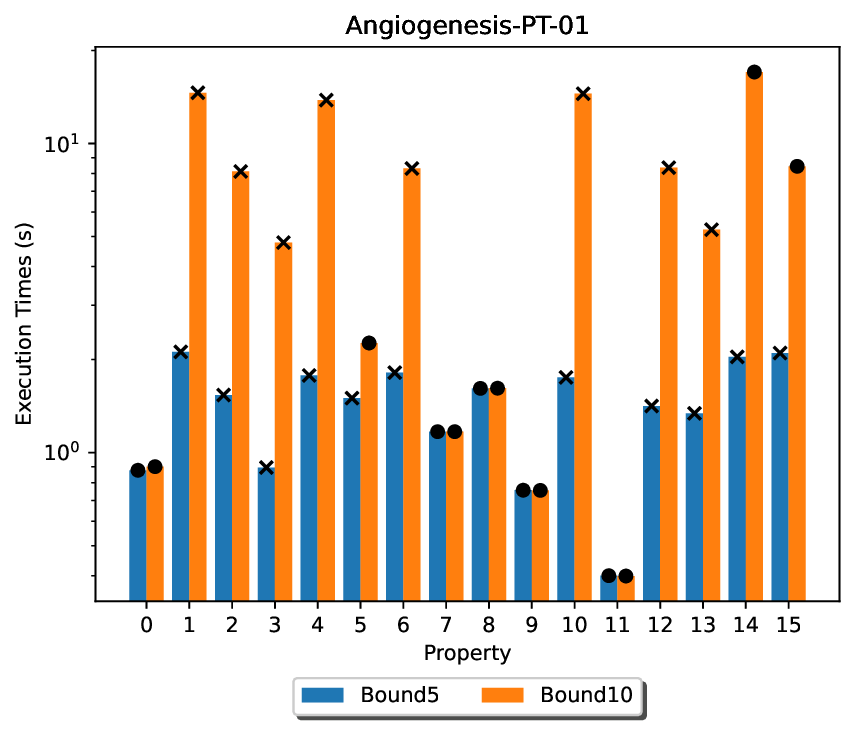}}
    \caption{Verification of FireabilityCardinality Properties with bound 5 and 10}
    \label{FCAngioPlot}

\end{figure}

\subsection{Experiment 2: Verification of FireabilityCardinality properties}\label{exp:compv1v2}

In this set of experiments, we verify a set of $16$ synthetic properties each on $20$ benchmarks from MCC. We introduced the set of FireabilityCardinality properties as a combination of the LTLFireability and LTLCardinality properties. For instance, $G (F(t_0) \land F (\#p_1 >999) $, which is a conjunction of the temporal property describing the firing of transition $t_0$ and the counting property with place $p_1$.

On DCModelChecker 2.0, for each of these $20$ benchmarks, we verified a set of $16$ FireabilityCardinality properties for each model, totalling 320 properties and 1280 runs with a timeout of $3600$ seconds each. The execution times (seconds) are plotted as separate bar graphs for bounds 5 and 10. As a sample, 16 properties were verified for the model Angiogenesis-PT-01 as shown in Fig~\ref{FCAngioPlot}. Each instance represents a FireabilityCardinality property. If the property is false and the tool returns a counterexample trace, it is depicted by a $\bullet$ and the symbol $\times$ denotes that the tool returned unsatisfiable (indicating that the property may be false at a greater bound). In Fig~\ref{FCAngioPlot}, we can observe that for bound 5 (blue line), properties $14$ and $15$ were false. On increasing the bound to 10, we obtain a satisfiable answer with counterexample traces for the same properties as more search space is explored by the tool, which also takes more time. This trend is observed in all models in our experiment, when the bound is increased, new counterexamples may be found by tool. The detailed plots are in Appendix.~\ref{sec:app}.

\subsection{Experiment 3: Verification of Unbounded Petri nets}\label{exp:upn}
To quote Amat et al, ``It is difficult to find benchmarks for unbounded Petri nets''~\cite{AmatDH22}. The strength of DCModelChecker 2.0 lies in verifying unbounded Petri nets. Since, MCC benchmarks are bounded~\cite{AmatDH22}, we make use of the unbounded Petri nets given by~\cite{NAmatGH} and verify LTLFireability properties in comparison with DCModelChecker 1.0~\cite{Zen2DBMC}, ITS-Tools~\cite{ITSMCC} and TAPAAL~\cite{DavidJJJMS12}. The results of these experiments are in Table~\ref{table:unboundedpn}, and~\cite{Zen2DBMC}. Each property that was verified is false. In each case it was observed that DCModelChecker 2.0 and DCModelChecker 1.0 returned that the property is false and additionally gave a counterexample trace, whereas ITS-Tools and TAPAAL returned that the property is false. In Section~\ref{exp:upn}, we verified LTLFireability properties of unbounded Petri nets which were all found to be false.
The properties are described in Table~\ref{table:upnprop}. The detailed output is available in the artifact~\cite{Zen2DBMC}.
\begin{table}[ht]
    \centering
    \resizebox{0.7\columnwidth}{!}{

        \begin{tabular}{|c|c|c|c|c|}
            \hline
            \multirow{2}{*}{Model} & \multicolumn{4}{c|}{Execution Times(ms)}                                                              \\
            \cline{2-5}
                                   & DCModelChecker 1.0 (bound =5)            & DCModelChecker 2.0 (bound =5) & ITS-Tools & Tapaal         \\
            \hline
            Parity                 & \textbf{0.010}                           & 0.40                          & 0.933     & 0.02           \\
            \hline
            PGCD                   & \textbf{0.019}                           & 0.98                          & 1.487     & 0.07           \\
            \hline
            Process                & 0.033                                    & 0.85                          & 1.201     & \textbf{1e-05} \\
            \hline
            CryptoMiner            & 0.036                                    & 0.64                          & 0.957     & \textbf{7e-06} \\
            \hline
            Murphy                 & 0.031                                    & 0.83                          & 1.161     & \textbf{8e-06} \\
            \hline
        \end{tabular}
    }
    \caption{Results of Experiment 3}
    \label{table:unboundedpn}

\end{table}

\begin{table}
    \centering

    \begin{tabular}{|l|l|}
        \hline
        \textbf{Model} & \textbf{Property}                 \\
        \hline
        Parity         & $\neg(t_0 U t_1)$                 \\
        \hline
        PGCD           & $\neg GF(t_0 U t_1)$              \\
        \hline
        Process        & $\neg F(t_0 U t_1)$               \\
        \hline
        CryptoMiner    & $\neg F(\text{OB } U \text{ GH})$ \\
        \hline
        Murphy         & $\neg F(t_1 U t_4)$               \\
        \hline
    \end{tabular}
    \caption{Experiment 3: List of properties verified on unbounded nets}
    \label{table:upnprop}
\end{table}

\subsection{Experiment 4: Verification of unbounded PN synthetic benchmark}\label{exp:aps}

We have additionally created a synthetic benchmark and properties specified in the {\LC} and verified them using DCModelChecker2.0. The specifications are available in our artifact~\cite{Zen2DBMCv2}. Our experiments demonstrate that DCModelChecker 2.0 has an advantage over DCModelChecker 1.0 in being able to additionally verify invariants. And it can be added to the arsenal of BMC tools for Petri nets.

\subsection{Limitations}\label{ssec:limitationsltllia}
In Experiment 1, DCModelChecker1.0 gave false positives on 7 out of 352 instances and has a tool confidence of $0.90$.
During our testing phase, we observed that several of these errors were due to issues in translation of the formula.
In order to avoid such errors, in future, we plan to implement a more robust translation of the LTL formula into the format readable by our tool.

Currently, DCModelChecker 2.0 does not take the truly concurrent execution for every run.
In future, we would like to build a BMC tool that behaves truly concurrently in every step. This will also enable us to compare the interleaving and concurrent semantics.

DCModelChecker tools are not yet optimized for performance and lags behind the state of the art tools such as ITS-Tools and Tapaal in some categories.
It is to be noted that our tools additionally display a counterexample trace and detailed logs when the property fails. This accounts for a significant portion of execution times.
As part of future work, we would like to give an optimal encoding of nets.

\subsection{Related Petri net verification tools}
There are several tools such as KREACH~\cite{DixonL20}, Petrinizer~\cite{EsparzaLMMN14}, QCOVER~\cite{BlondinFHH16}, ICOVER~\cite{GeffroyLS18} verifying specific classes of properties like reachability and coverability of Petri nets. Recently, in~\cite{AmatBD21,AmatDH22} the generalized reachability of Petri net is encoded into a BMC problem and solved using SMT solvers. However, BMC for unbounded Petri nets with concurrency and LTL properties while leveraging the power of SMT Solvers remained unexplored until now. DCModelChecker bridges this gap.

The MCC~\cite{MCC} has attracted many tools for formal verification of concurrent systems that use a portfolio approach. In particular, we looked at ITS-Tools~\cite{ITSMCC} which implements structural reduction techniques~\cite{Mieg20} and uses a layer of SMT when solving LTL. Moreover, it does not perform BMC. While ITS-Tools is much faster on bounded nets (MCC benchmarks) it gives only binary answers (sat or unsat). From our experiments, on a subset of unbounded nets, while TAPAAL is faster, it also gives binary answers. On the contrary, our tool preserves the original Petri net structure and additionally gives a counterexample trace which is useful in bug finding in systems.

\section{Conclusions and Future Work}\label{sec:concl}

In this work, we introduced the technique of two dimensional bounded model checking and showed experimental results where our BMC tool has sound results an also where it competes with the state of the art tools. It is noteworthy that our tool can additionally generate the counterexample trace, when the property fails. This is a crucial piece of information to troubleshoot the verification process. As part of future work, we intend to make use of this trace, to automatically refine the model and specification along the lines of the CEGAR approach~\cite{CEGARWolf12,HajduVBM14}.

\bibliographystyle{splncs04}
\bibliography{references}

\begin{thebibliography}{10}
\providecommand{\url}[1]{\texttt{#1}}
\providecommand{\urlprefix}{URL }
\providecommand{\doi}[1]{https://doi.org/#1}

\bibitem{AbdullaIN00}
Abdulla, P.A., Iyer, S.P., Nyl{\'{e}}n, A.: Unfoldings of unbounded {P}etri
  nets. In: Computer Aided Verification, 12th International Conference, {CAV}
  2000, Chicago, IL, USA, July 15-19, 2000, Proceedings. Lecture Notes in
  Computer Science, vol.~1855, pp. 495--507. Springer (2000).
  \doi{10.1007/10722167\_37}, \url{https://doi.org/10.1007/10722167\_37}

\bibitem{Abdulla04}
Abdulla, P.A., Iyer, S.P., Nyl\'{e}n, A.: {SAT}-solving the coverability
  problem for {P}etri nets. Form. Methods Syst. Des.  \textbf{24}(1),  25–43
  (Jan 2004). \doi{10.1023/B:FORM.0000004786.30007.f8}

\bibitem{AbdullaJ01}
Abdulla, P.A., Jonsson, B.: Ensuring completeness of symbolic verification
  methods for infinite-state systems. Theor. Comput. Sci.  \textbf{256}(1-2),
  145--167 (2001). \doi{10.1016/S0304-3975(00)00105-5},
  \url{https://doi.org/10.1016/S0304-3975(00)00105-5}

\bibitem{AbdullaJRS06}
Abdulla, P.A., Jonsson, B., Rezine, A., Saksena, M.: Proving liveness by
  backwards reachability. In: {CONCUR} 2006 - Concurrency Theory, 17th
  International Conference, {CONCUR} 2006, Bonn, Germany, August 27-30, 2006,
  Proceedings. Lecture Notes in Computer Science, vol.~4137, pp. 95--109.
  Springer (2006), \url{https://doi.org/10.1007/11817949\_7}

\bibitem{AmatBD21}
Amat, N., Berthomieu, B., Dal{-}Zilio, S.: On the combination of polyhedral
  abstraction and smt-based model checking for {Petri} nets. In: {PETRI}
  {NETS}. pp. 164--185. LNCS, Springer (2021).
  \doi{10.1007/978-3-030-76983-3\_9}

\bibitem{AmatDH22}
Amat, N., Dal{-}Zilio, S., Hujsa, T.: Property directed reachability for
  generalized {P}etri nets. In: Tools and Algorithms for the Construction and
  Analysis of Systems - 28th International Conference, {TACAS} 2022, Held as
  Part of the European Joint Conferences on Theory and Practice of Software,
  {ETAPS} 2022, Munich, Germany, April 2-7, 2022, Proceedings, Part {I}.
  Lecture Notes in Computer Science, vol. 13243, pp. 505--523. Springer (2022).
  \doi{10.1007/978-3-030-99524-9\_28},
  \url{https://doi.org/10.1007/978-3-030-99524-9\_28}

\bibitem{BarrettCDHJKRT11}
Barrett, C.W., Conway, C.L., Deters, M., Hadarean, L., Jovanovic, D., King, T.,
  Reynolds, A., Tinelli, C.: {CVC4}. In: Computer Aided Verification - 23rd
  International Conference, {CAV} 2011, Snowbird, UT, USA, July 14-20, 2011.
  Proceedings. Lecture Notes in Computer Science, vol.~6806, pp. 171--177.
  Springer (2011). \doi{10.1007/978-3-642-22110-1\_14},
  \url{https://doi.org/10.1007/978-3-642-22110-1\_14}

\bibitem{HandbookSMTBarrettT18}
Barrett, C.W., Tinelli, C.: Satisfiability modulo theories. In: Clarke, E.M.,
  Henzinger, T.A., Veith, H., Bloem, R. (eds.) Handbook of Model Checking, pp.
  305--343. Springer (2018). \doi{10.1007/978-3-319-10575-8\_11},
  \url{https://doi.org/10.1007/978-3-319-10575-8\_11}

\bibitem{BerthomieuV06}
Berthomieu, B., Vernadat, F.: Time {Petri} nets analysis with {TINA}. In: Third
  International Conference on the Quantitative Evaluation of Systems {(QEST}
  2006), 11-14 September 2006, Riverside, California, {USA}. pp. 123--124.
  {IEEE} Computer Society (2006). \doi{10.1109/QEST.2006.56},
  \url{https://doi.org/10.1109/QEST.2006.56}

\bibitem{BiereCCSZ03}
Biere, A., Cimatti, A., Clarke, E.M., Strichman, O., Zhu, Y.: Bounded model
  checking. Adv. Comput. pp. 117--148 (2003).
  \doi{10.1016/S0065-2458(03)58003-2}

\bibitem{BiereTACAS99}
Biere, A., Cimatti, A., Clarke, E.M., Zhu, Y.: Symbolic model checking without
  bdds. In: {TACAS}. pp. 193--207. LNCS, Springer (1999).
  \doi{10.1007/3-540-49059-0\_14}

\bibitem{BlondinFHH16}
Blondin, M., Finkel, A., Haase, C., Haddad, S.: Approaching the coverability
  problem continuously. In: {TACAS}. pp. 480--496. LNCS, Springer (2016).
  \doi{10.1007/978-3-662-49674-9\_28}

\bibitem{ChengEP95}
Cheng, A., Esparza, J., Palsberg, J.: Complexity results for 1-safe nets.
  Theor. Comput. Sci.  \textbf{147}(1{\&}2),  117--136 (1995).
  \doi{10.1016/0304-3975(94)00231-7}

\bibitem{ClarkeKNZ11}
Clarke, E.M., Klieber, W., Nov{\'{a}}cek, M., Zuliani, P.: Model checking and
  the state explosion problem. In: Tools for Practical Software Verification,
  LASER, International Summer School 2011, Elba Island, Italy, Revised Tutorial
  Lectures. Lecture Notes in Computer Science, vol.~7682, pp. 1--30. Springer
  (2011), \url{https://doi.org/10.1007/978-3-642-35746-6\_1}

\bibitem{CzerwinskiLLLM21}
Czerwinski, W., Lasota, S., Lazic, R., Leroux, J., Mazowiecki, F.: The
  reachability problem for {Petri} nets is not elementary. J. {ACM}
  \textbf{68}(1),  7:1--7:28 (2021), \url{https://doi.org/10.1145/3422822}

\bibitem{DavidJJJMS12}
David, A., Jacobsen, L., Jacobsen, M., J{\o}rgensen, K.Y., M{\o}ller, M.H.,
  Srba, J.: {TAPAAL} 2.0: Integrated development environment for timed-arc
  {Petri} nets. In: TACAS. LNCS, vol.~7214, pp. 492--497. Springer (2012).
  \doi{10.1007/978-3-642-28756-5\_36}

\bibitem{DixonL20}
Dixon, A., Lazic, R.: Kreach: {A} tool for reachability in {Petri} nets. In:
  {TACAS}. pp. 405--412. LNCS, Springer (2020).
  \doi{10.1007/978-3-030-45190-5\_22}

\bibitem{EsparzaH01}
Esparza, J., Heljanko, K.: Implementing {LTL} model checking with net
  unfoldings. In: Dwyer, M.B. (ed.) Model Checking Software, 8th International
  {SPIN} Workshop, Toronto, Canada, May 19-20, 2001, Proceedings. Lecture Notes
  in Computer Science, vol.~2057, pp. 37--56. Springer (2001).
  \doi{10.1007/3-540-45139-0\_4}

\bibitem{EsparzaLMMN14}
Esparza, J., Ledesma{-}Garza, R., Majumdar, R., Meyer, P.J., Niksic, F.: An
  smt-based approach to coverability analysis. In: CAV. pp. 603--619. LNCS,
  Springer (2014). \doi{10.1007/978-3-319-08867-9\_40}

\bibitem{GeffroyLS18}
Geffroy, T., Leroux, J., Sutre, G.: Occam's razor applied to the {Petri} net
  coverability problem. Theor. Comput. Sci.  (2018).
  \doi{10.1016/j.tcs.2018.04.014}

\bibitem{Hack74}
Hack, M.: The recursive equivalence of the reachability problem and the
  liveness problem for {Petri} nets and vector addition systems. In: 15th
  Annual Symposium on Switching and Automata Theory. pp. 156--164. {IEEE}
  Computer Society (1974). \doi{10.1109/SWAT.1974.28}

\bibitem{HajduVBM14}
Hajdu, {\'{A}}., V{\"{o}}r{\"{o}}s, A., Bartha, T., M{\'{a}}rtonka, Z.:
  Extensions to the {CEGAR} approach on petri nets. Acta Cybern.
  \textbf{21}(3),  401--417 (2014). \doi{10.14232/ACTACYB.21.3.2014.8},
  \url{https://doi.org/10.14232/actacyb.21.3.2014.8}

\bibitem{HillahKPT10}
Hillah, L., Kordon, F., Petrucci, L., Tr{\`{e}}ves, N.: {PNML} framework: An
  extendable reference implementation of the {Petri} net markup language. In:
  {PETRI} {NETS}. pp. 318--327. LNCS, Springer (2010).
  \doi{10.1007/978-3-642-13675-7\_20}

\bibitem{MCC}
Kordon, F., Bouvier, P., Garavel, H., Hillah, L.M., Hulin-Hubard, F., Amat.,
  N., Amparore, E., Berthomieu, B., Biswal, S., Donatelli, D., Galla, F., ,
  {Dal Zilio}, S., Jensen, P., He, C., {Le Botlan}, D., Li, S., , Srba, J.,
  Thierry-Mieg, ., Walner, A., Wolf, K.: {Complete Results for the 2020 Edition
  of the Model Checking Contest}. {http://mcc.lip6.fr/2021/results.php} (June
  2021)

\bibitem{CompResultMCC22}
{Results of Model Checking Contest}. {https://mcc.lip6.fr/results.php}
  (September 2022)

\bibitem{KenUnfolding92}
McMillan, K.L.: Using unfoldings to avoid the state explosion problem in the
  verification of asynchronous circuits. In: Proceedings of the Fourth
  International Workshop on Computer Aided Verification. p. 164–177. CAV '92,
  Springer-Verlag, Berlin, Heidelberg (1992)

\bibitem{MouraB08}
de~Moura, L.M., Bj{\o}rner, N.S.: {Z3:} an efficient {SMT} solver. In: {TACAS}.
  pp. 337--340. LNCS, Springer (2008). \doi{10.1007/978-3-540-78800-3\_24}

\bibitem{murata89}
Murata, T.: Petri nets: Properties, analysis and applications. IEEE
  \textbf{77}(4),  541--580 (1989)

\bibitem{NAmatGH}
Nicolas, A.: {Benchmarks of Unbounded Petri Nets}.
  {https://github.com/nicolasAmat/SMPT} (September 2022)

\bibitem{ANTLRParrF11}
Parr, T., Fisher, K.: Ll(*): the foundation of the {ANTLR} parser generator.
  In: {PLDI}. pp. 425--436 (2011). \doi{10.1145/1993498.1993548}

\bibitem{petri1962kommunikation}
Petri, C.A.: {Kommunikation mit Automaten}. Dissertation, Schriften des IIM~2,
  Rheinisch-Westf{\"a}lisches Institut f{\"u}r Instrumentelle Mathematik an der
  Universit{\"a}t Bonn, Bonn (1962)

\bibitem{FigPrince2022}
Phawade, R., Prince, T., Sheerazuddin, S.: {Artifact and instructions to
  generate experimental results for Two Dimensional Bounded Model Checking of
  Unbounded Client-Server Systems v3} (4 2022).
  \doi{10.6084/m9.figshare.19611477.v3},
  \url{https://figshare.com/articles/software/Untitled_Item/19611477}

\bibitem{arxiv2dbmc}
Phawade, R., Prince, T., Sheerazuddin, S.: Bounded model checking for unbounded
  client server systems (2022), \url{https://arxiv.org/abs/2209.05879}

\bibitem{Zen2DBMCv2}
Phawade, R., Prince, T., Sheerazuddin, S.: Dcmodelchecker 2.0: A bmc tool for
  unbounded pn (2022). \doi{10.5281/zenodo.7198484}

\bibitem{Zen2DBMC}
Phawade, R., Prince, T., Sheerazuddin, S.: Dcmodelchecker1.0 tool for
  verification of unbounded {Petri} nets (2022). \doi{10.5281/zenodo.7084996}

\bibitem{P77}
Pnueli, A.: The temporal logic of programs. In: FOCS. pp. 46--57 (1977).
  \doi{10.1109/SFCS.1977.32}

\bibitem{Rackoff78}
Rackoff, C.: The covering and boundedness problems for vector addition systems.
  Theoretical Computer Science  \textbf{6}(2),  223--231 (1978).
  \doi{https://doi.org/10.1016/0304-3975(78)90036-1}

\bibitem{RulesMCC22}
{Model Checking Contest 2022 Rules}. {https://mcc.lip6.fr/pdf/rules.pdf}
  (September 2022)

\bibitem{ITSMCC}
Thierry{-}Mieg, Y.: Symbolic model-checking using its-tools. In: {TACAS}. pp.
  231--237. LNCS, Springer (2015). \doi{10.1007/978-3-662-46681-0\_20}

\bibitem{Mieg20}
Thierry{-}Mieg, Y.: Structural reductions revisited. In: {PETRI} {NETS}. pp.
  303--323. LNCS, Springer (2020). \doi{10.1007/978-3-030-51831-8\_15}

\bibitem{VardiW86}
Vardi, M.Y., Wolper, P.: An automata-theoretic approach to automatic program
  verification (preliminary report). In: Proceedings of the Symposium on Logic
  in Computer Science {(LICS} '86), Cambridge, Massachusetts, USA, June 16-18,
  1986. pp. 332--344. {IEEE} Computer Society (1986)

\bibitem{CEGARWolf12}
Wimmel, H., Wolf, K.: Applying {CEGAR} to the petri net state equation. Log.
  Methods Comput. Sci.  \textbf{8}(3) (2012). \doi{10.2168/LMCS-8(3:27)2012},
  \url{https://doi.org/10.2168/LMCS-8(3:27)2012}

\bibitem{PNWolfgang}
Zahoransky, R.M., Holderer, J., Lange, A., Brenig, C.: Process analysis as
  first step towards automated business security. In: {ECIS}. p. Research Paper
  46 (2016), \url{http://aisel.aisnet.org/ecis2016\_rp/46}

\end{thebibliography}
\pagebreak
\appendix
\section{Appendix: Details of Experiment 2: Verification of FireabilityCardinality properties}\label{sec:app}
In Sec.~\ref{exp:compv1v2}, Experiment 2 was discussed. We verified FireabilityCardinality properties for Petri net benchmarks using our tool DCModelChecker2.0~\cite{Zen2DBMCv2}. The plots for the same are given below:

\begin{table}[]
  \begin{tabular}{cc}
    \includegraphics[width=.5\linewidth]{Angiogenesis-PT-01_FCPlot.eps} &
    \includegraphics[width=.5\linewidth]{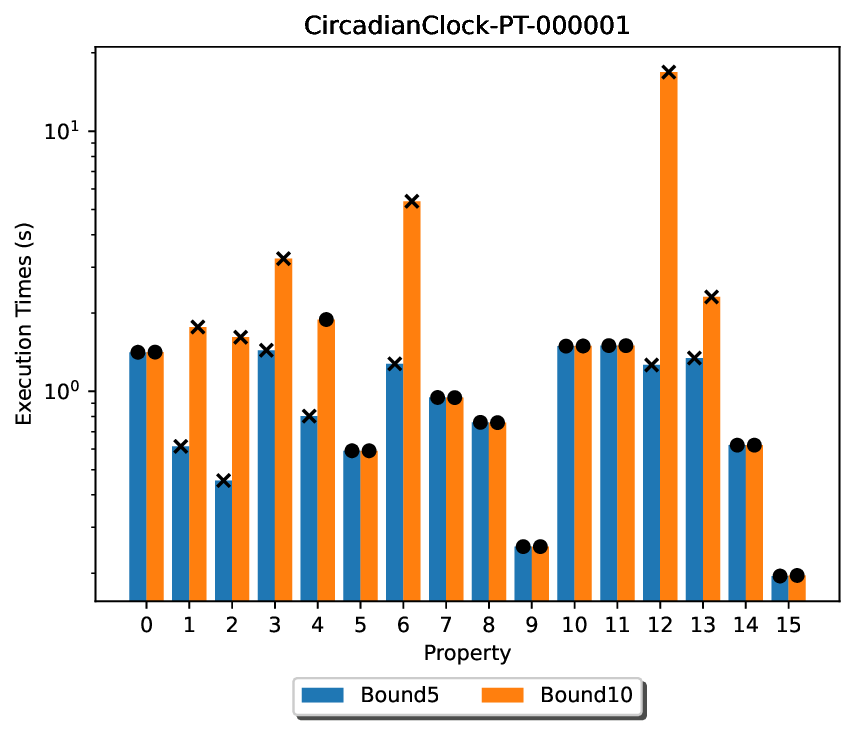} \\
  \end{tabular}
  \begin{tabular}{cc}
    \includegraphics[width=.5\linewidth]{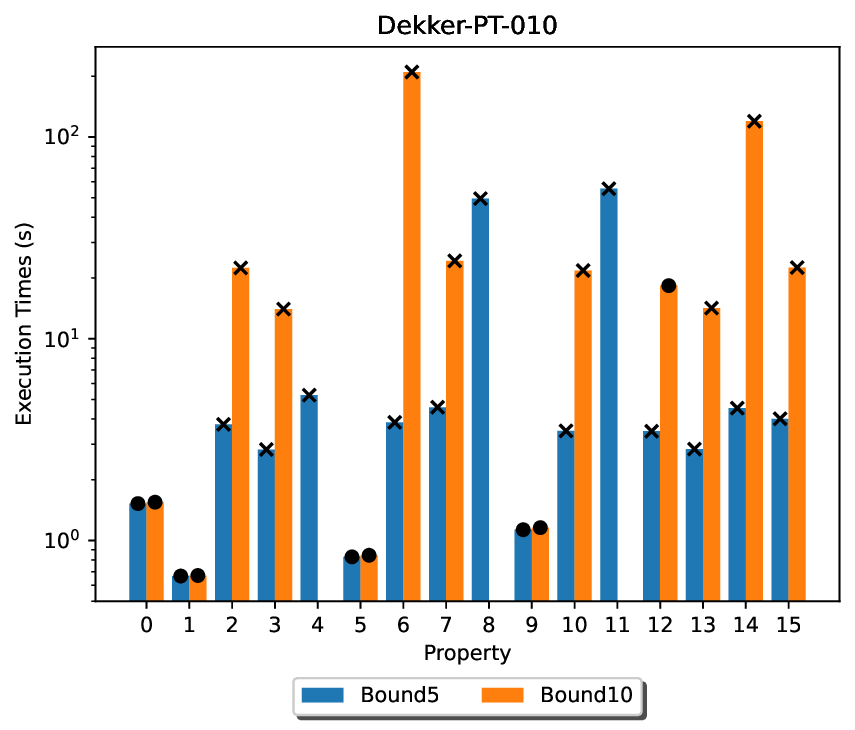} &
    \includegraphics[width=.5\linewidth]{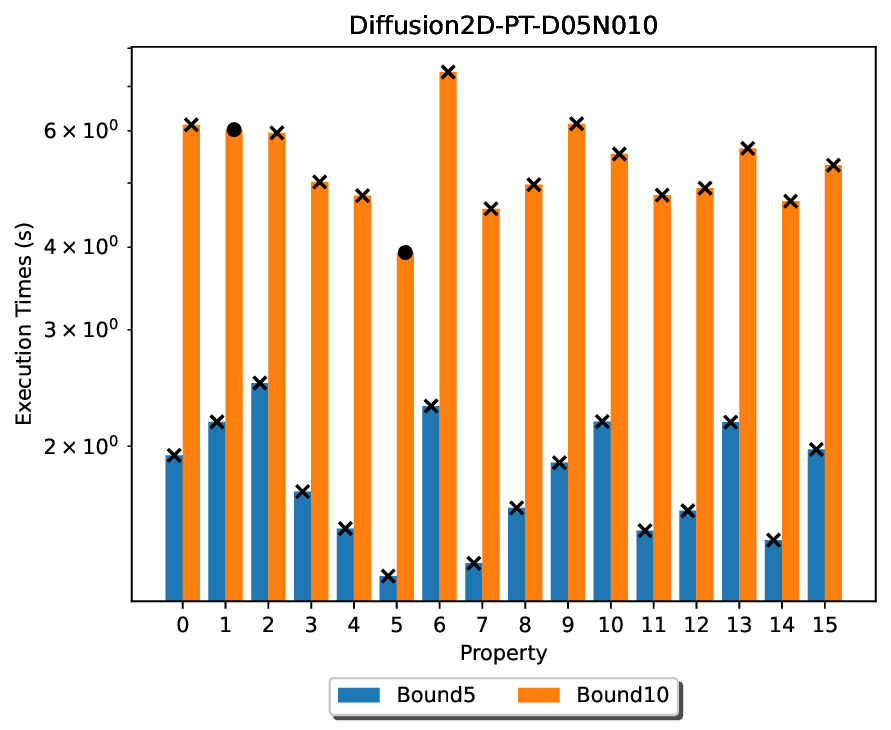} \\
  \end{tabular}
\end{table}
\begin{table}[]
  \begin{tabular}{cc}
    \includegraphics[width=.5\linewidth]{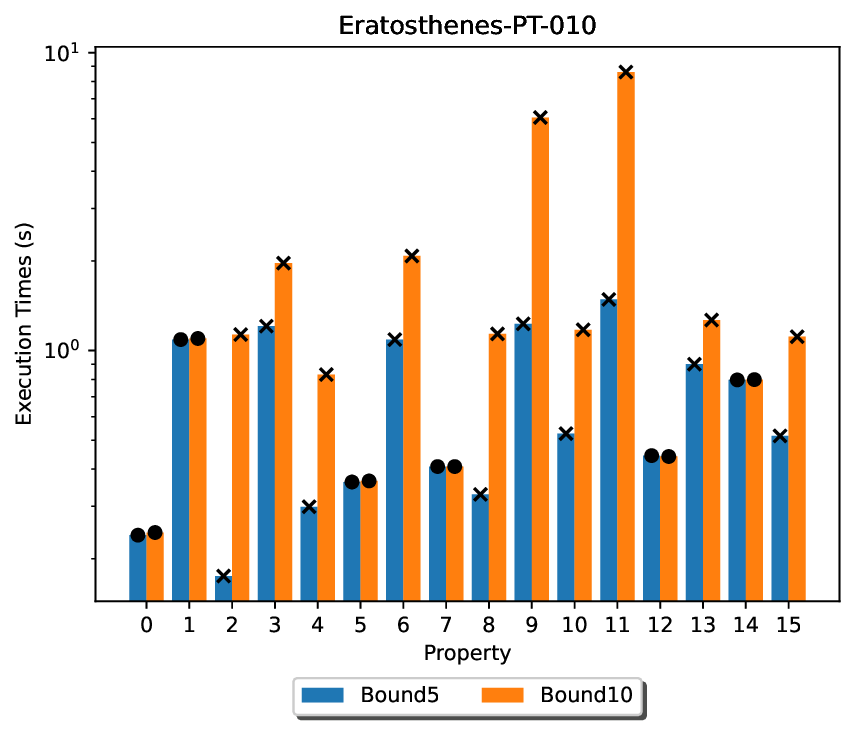} &
    \includegraphics[width=.5\linewidth]{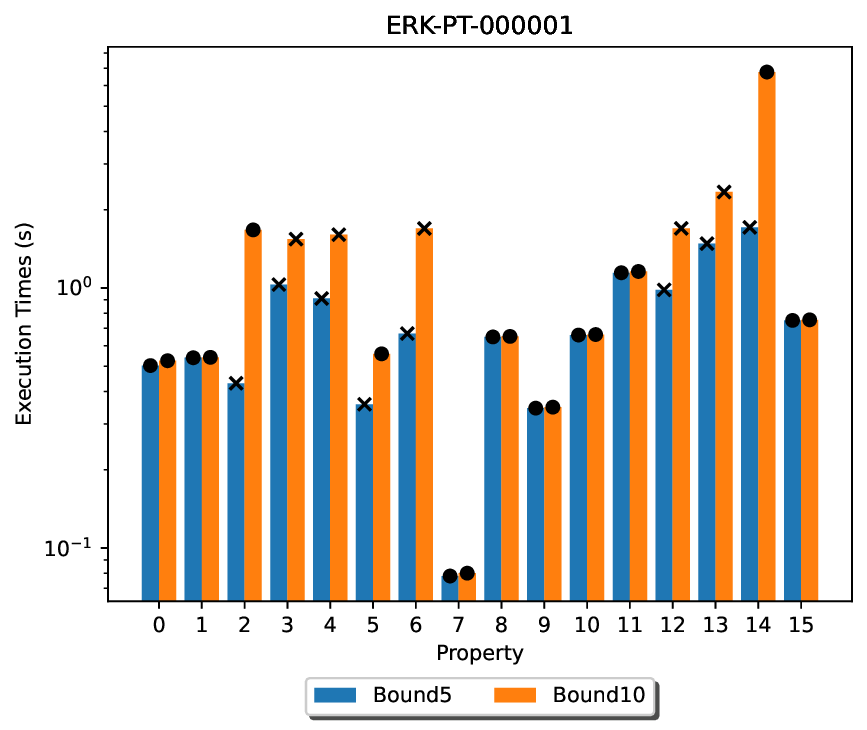}         \\
  \end{tabular}
  \begin{tabular}{cc}
    \includegraphics[width=.5\linewidth]{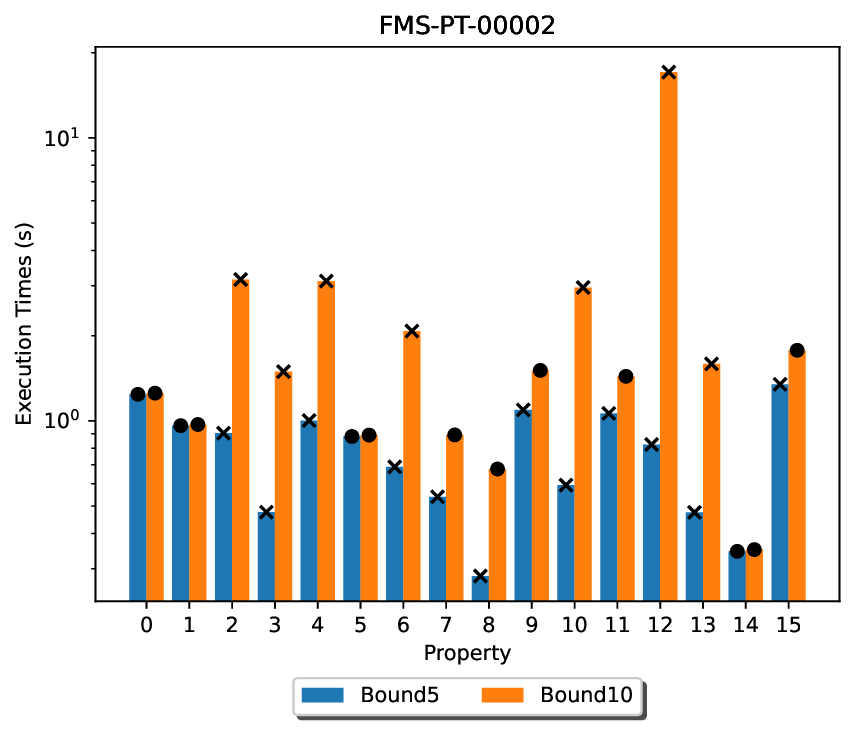} &
    \includegraphics[width=.5\linewidth]{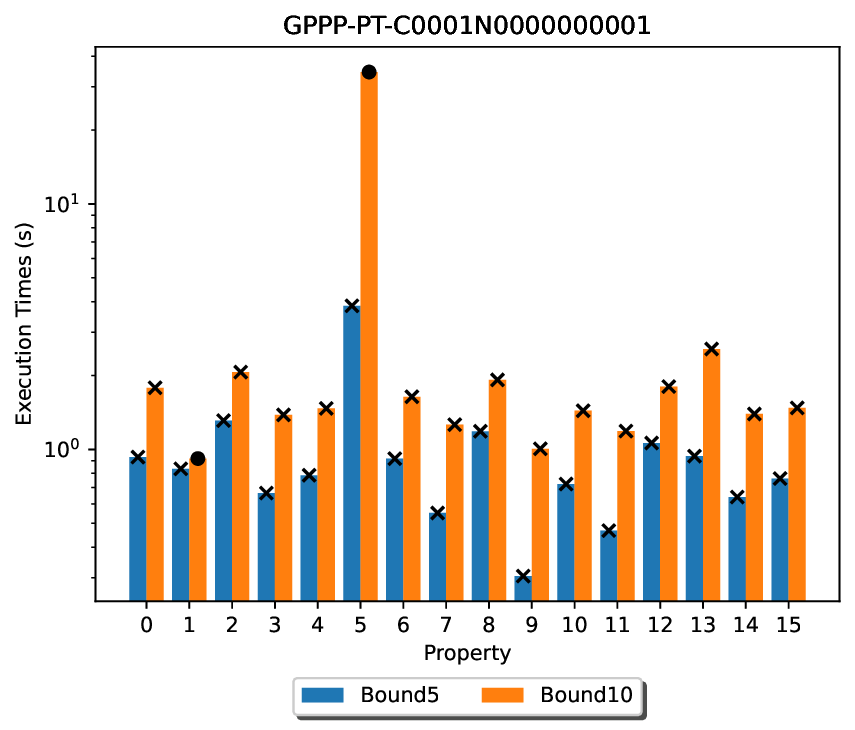} \\
  \end{tabular}
  \begin{tabular}{cc}
    \includegraphics[width=.5\linewidth]{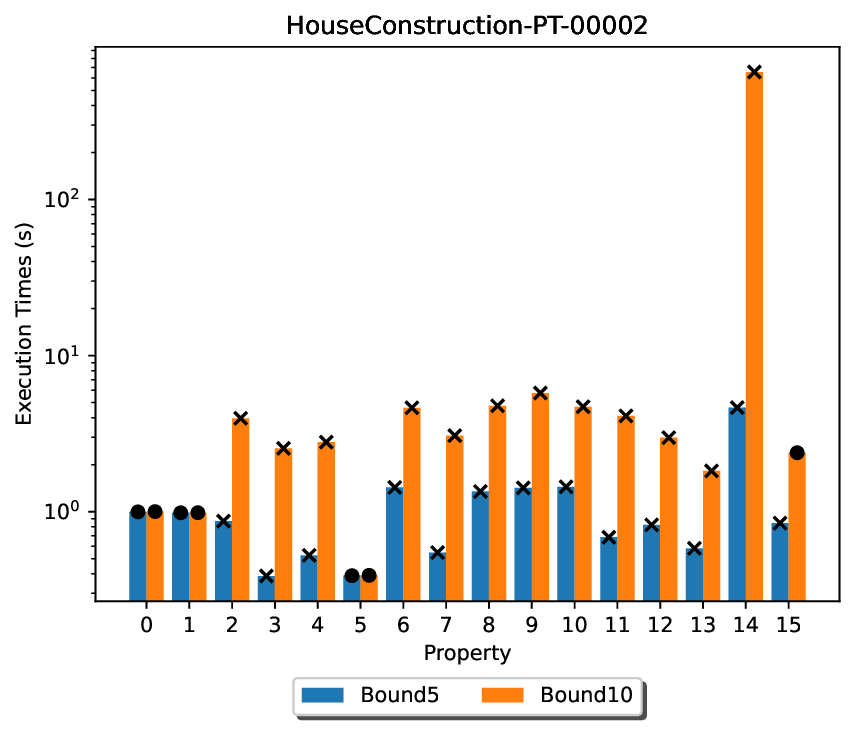} &
    \includegraphics[width=.5\linewidth]{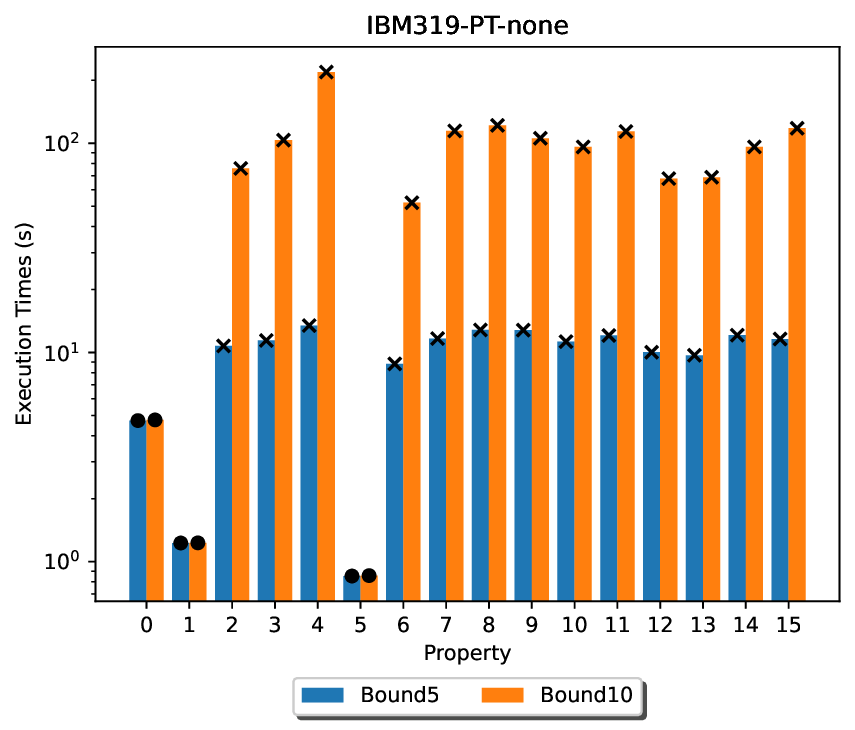}               \\
  \end{tabular}
\end{table}
\begin{table}[]
  \begin{tabular}{cc}
    \includegraphics[width=.5\linewidth]{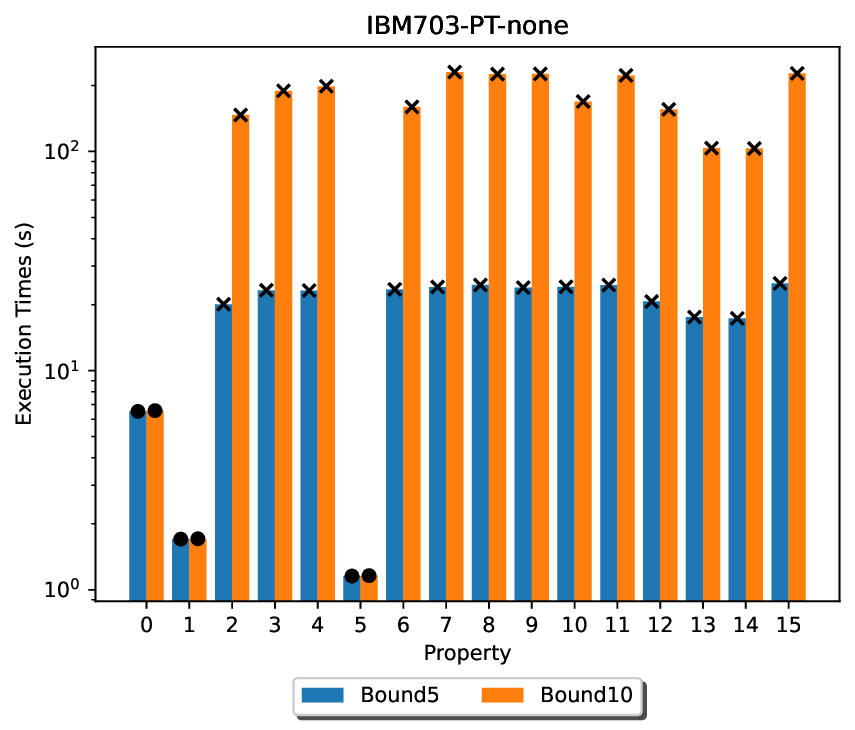} &
    \includegraphics[width=.5\linewidth]{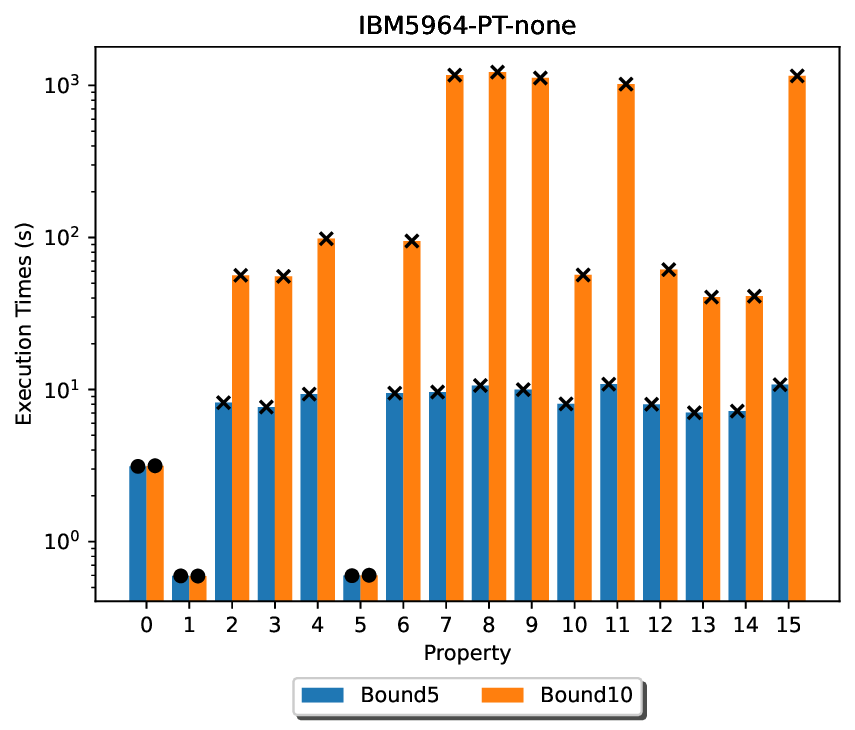}  \\
  \end{tabular}
  \begin{tabular}{cc}
    \includegraphics[width=.5\linewidth]{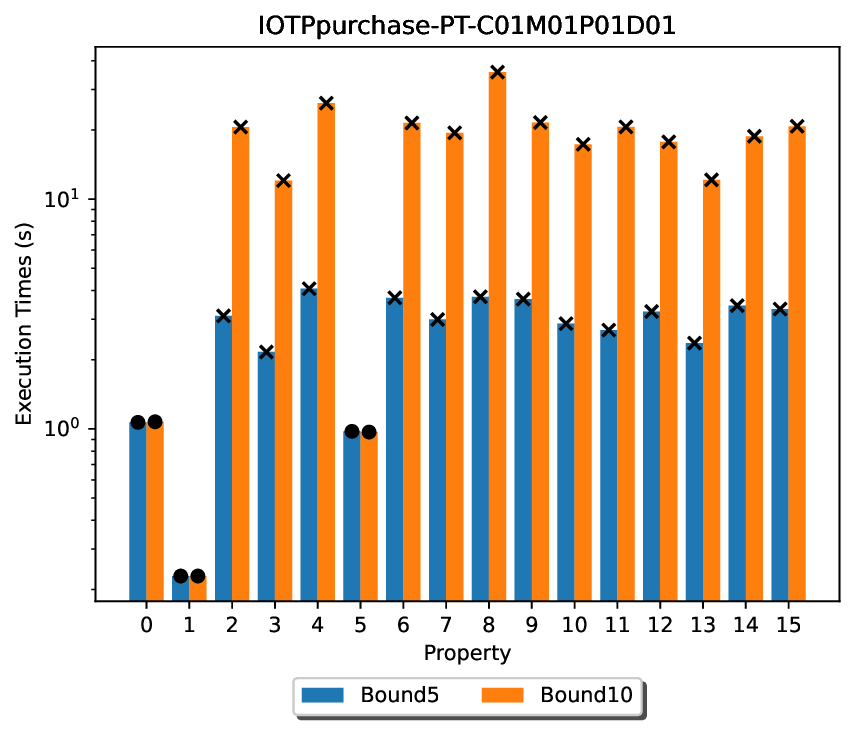} &
    \includegraphics[width=.5\linewidth]{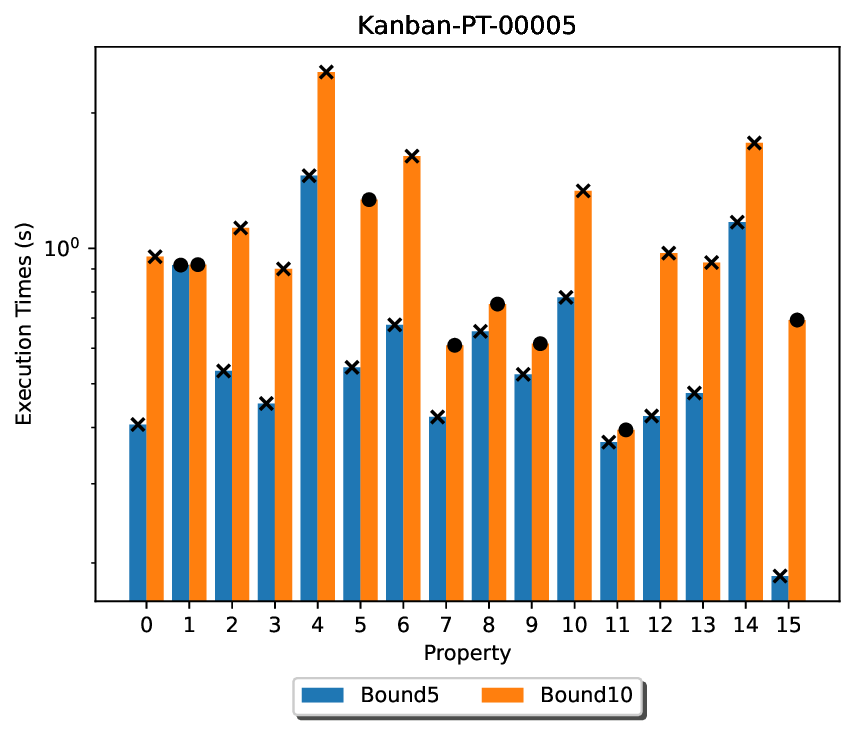}                \\
  \end{tabular}
\end{table}
\begin{table}[]
  \begin{tabular}{cc}
    \includegraphics[width=.5\linewidth]{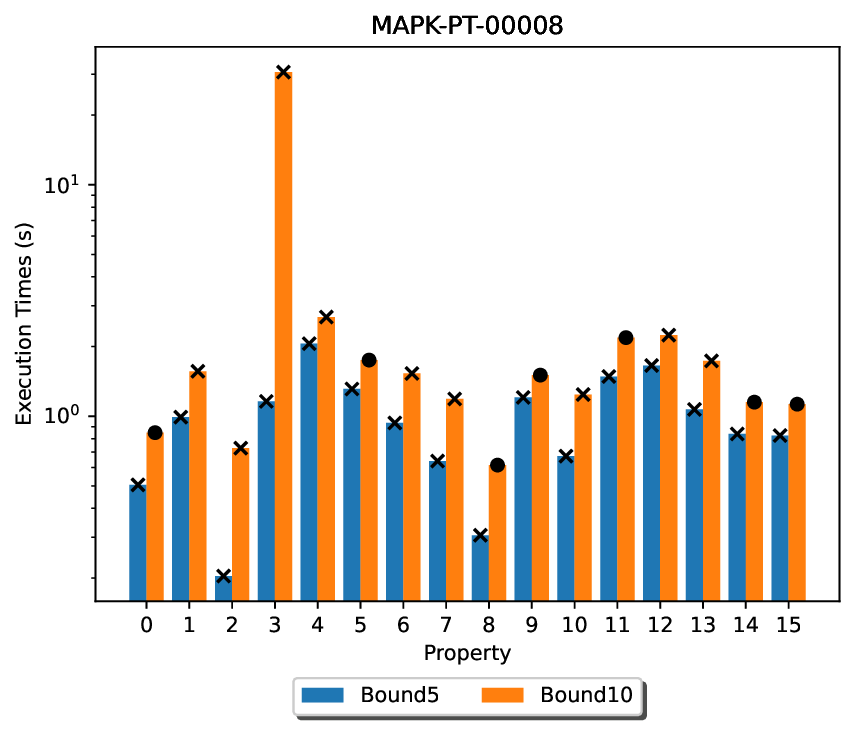} &
    \includegraphics[width=.5\linewidth]{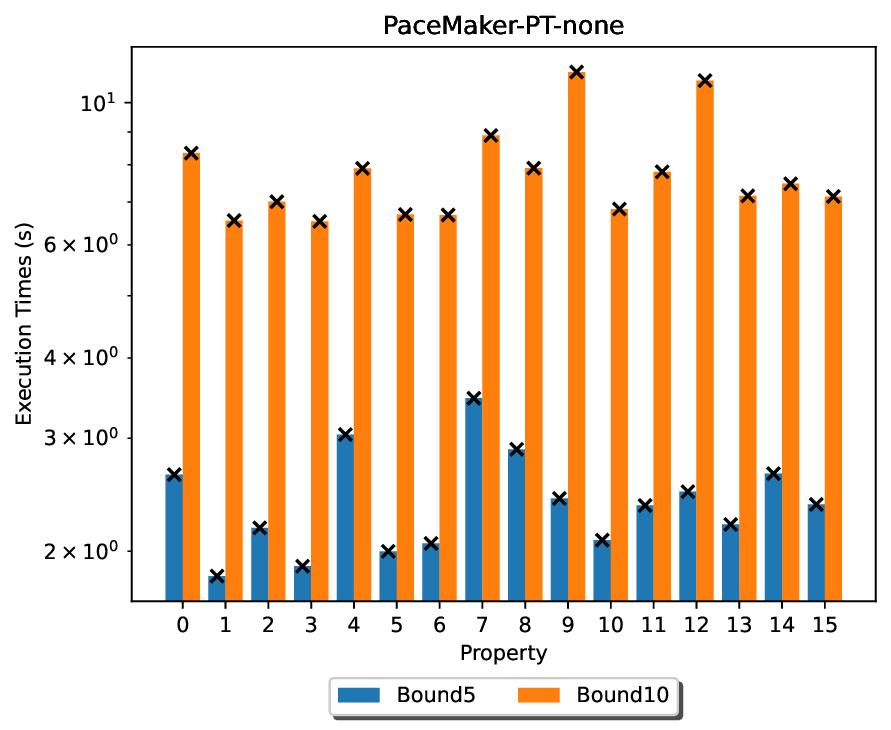} \\
  \end{tabular}
  \begin{tabular}{cc}
    \includegraphics[width=.5\linewidth]{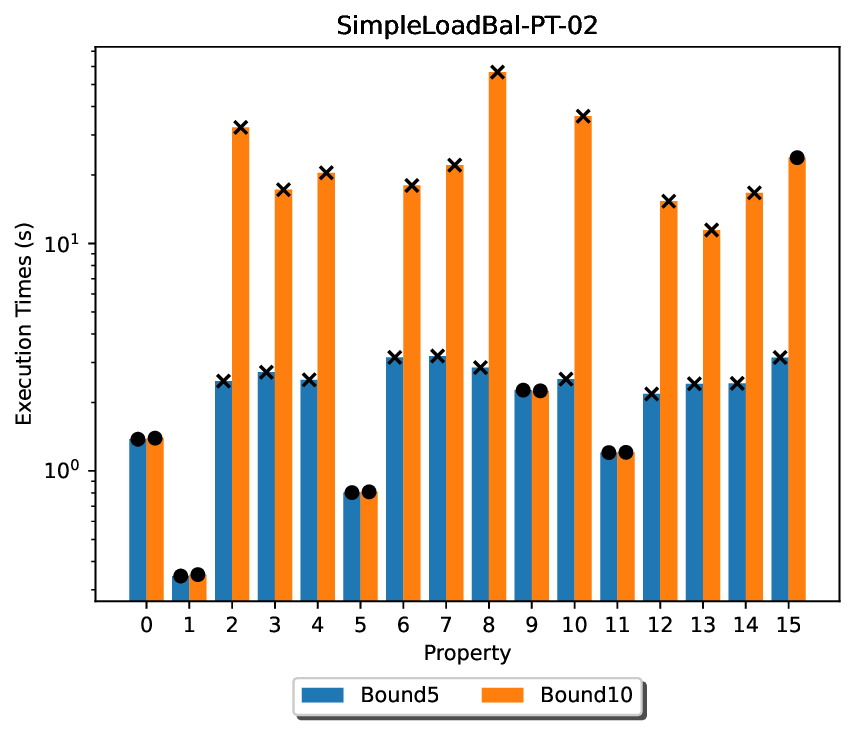} &
    \includegraphics[width=.5\linewidth]{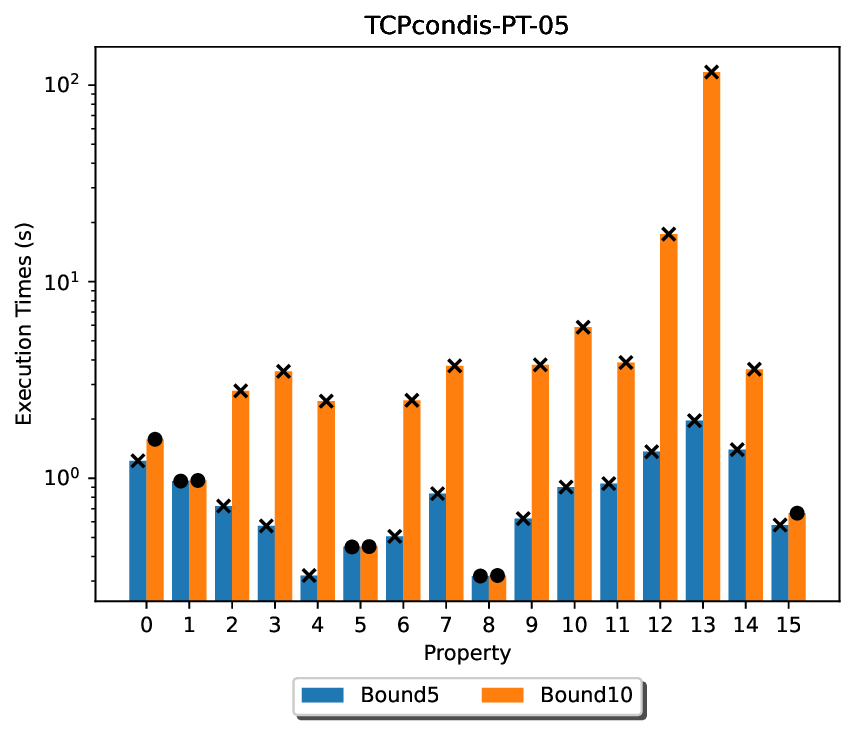}       \\
  \end{tabular}
  \begin{tabular}{cc}
    \includegraphics[width=.5\linewidth]{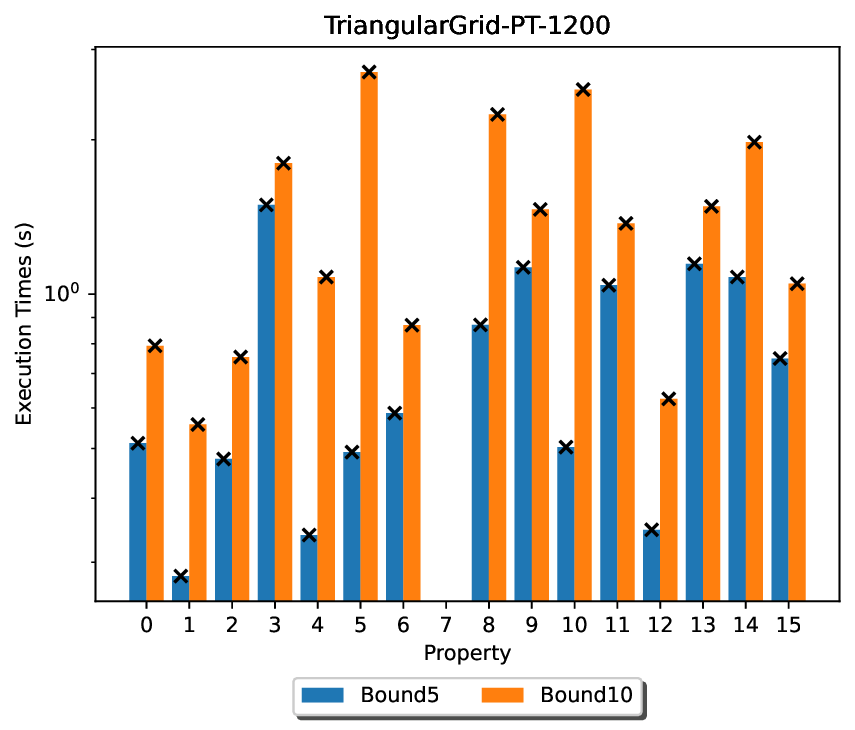} &
    \includegraphics[width=.5\linewidth]{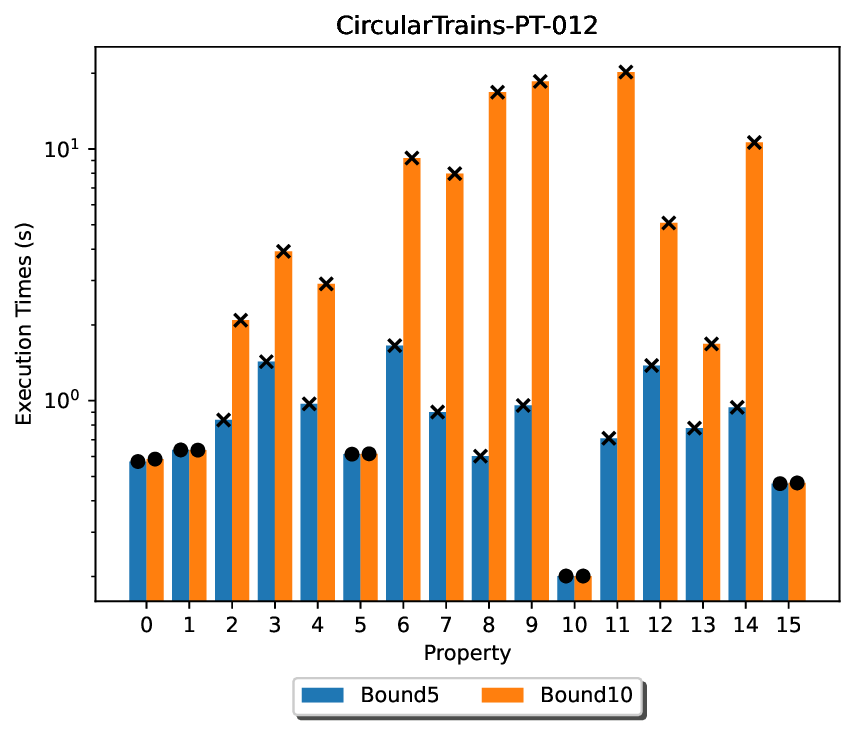}    \\
  \end{tabular}
\end{table}
\end{document}